\documentclass[journal,comsoc]{IEEEtran}
\usepackage{amsmath}
\usepackage{booktabs}  %  引入三线表宏包  
\usepackage{graphicx}
\usepackage{subfigure}
\usepackage[cmintegrals]{newtxmath}
\usepackage[T1]{fontenc}% optional T1 font encoding
\usepackage{epstopdf}
\usepackage{multirow}
\usepackage[figuresright]{rotating} % 支持表格横置
\usepackage{tabularx} % 支持表格自动换行
\usepackage{longtable} % 长表格
\usepackage{supertabular} % 表格自动分页
\usepackage{cite}
\usepackage{footnote}
\usepackage{tablefootnote}
\usepackage[colorlinks,
            linkcolor=blue,       %%修改此处为你想要的颜色
            anchorcolor=blue,  %%修改此处为你想要的颜色
            citecolor=blue,        %%修改此处为你想要的颜色,例如修改blue为red
            ]{hyperref}
\usepackage[numbers,sort&compress]{natbib} %将参考文献从[6,7,8,9]变成[6-9]
\usepackage{bbding}
\usepackage{array}
\usepackage{float}
\usepackage{makecell}
\usepackage{setspace}
\usepackage[table,xcdraw]{xcolor}
\usepackage{threeparttable} %给表格加脚注

\usepackage{url}

\begin{document}
\title{Machine Learning Empowered Intelligent Data Center Networking: A Survey}

\author{Bo~Li,~\IEEEmembership{Student Member,~IEEE,}
    Ting~Wang,~\IEEEmembership{Senior Member,~IEEE,}
    Peng~Yang,~\IEEEmembership{Student Member,~IEEE,}
    Mingsong~Chen,~\IEEEmembership{Senior Member,~IEEE,}
    Shui~Yu,~\IEEEmembership{Senior Member,~IEEE,}
    % and~Khaled~B. Letaief,~\IEEEmembership{Fellow,~IEEE}% <-this % stops a space
    and~Mounir~Hamdi~\IEEEmembership{Fellow,~IEEE}
    \thanks{(\textit{Corresponding author: Ting Wang.})}
    \thanks{Bo Li, Ting Wang, Peng Yang and Mingsong Chen are with the Software Engineering Institute, Shanghai Key Laboratory of Trustworthy Computing, East China Normal University, Shanghai, China (e-mail: 51205902130@stu.ecnu.edu.cn; twang@sei.ecnu.edu.cn; 51205902030@stu.ecnu.edu.cn; mschen@sei.ecnu.edu.cn).}
    \thanks{Shui Yu is with the School of Computer Science, University of Technology Sydney, Australia. (email: shui.yu@uts.edu.au).}
    % \thanks{Khaled B. Letaief is with Department of Electronic and Computer Engineering, Hong Kong University of Science and Technology, Hong Kong, China (email: eekhaled@ust.hk).}
    \thanks{Mounir Hamdi is with the College of Science and Engineering, Hamad Bin Khalifa University, Qatar (e-mail: mhamdi@hbku.edu.qa)}
}


\markboth{}%
{Shell \MakeLowercase{\textit{et al.}}: Bare Demo of IEEEtran.cls for IEEE Communications Society Journals}



\maketitle

\begin{abstract}
    To support the needs of ever-growing cloud-based services, the number of servers and network devices in data centers is increasing exponentially, which in turn results in high complexities and difficulties in network optimization. To address these challenges, both academia and industry turn to artificial intelligence technology to realize network intelligence. To this end, a considerable number of novel and creative machine learning-based (ML-based) research works have been put forward in recent few years. Nevertheless, there are still enormous challenges faced by the intelligent optimization of data center networks (DCNs), especially in the scenario of online real-time dynamic processing of massive heterogeneous services and traffic data. To best of our knowledge, there is a lack of systematic and original comprehensively investigations with in-depth analysis on intelligent DCN. To this end, in this paper, we comprehensively investigate the application of machine learning to data center networking, and provide a general overview and in-depth analysis of the recent works, covering flow prediction, flow classification, load balancing, resource management, routing optimization, and congestion control. In order to provide a multi-dimensional and multi-perspective comparison of various solutions, we design a quality assessment criteria called REBEL-3S to impartially measure the strengths and weaknesses of these research works. Moreover, we also present unique insights into the technology evolution of the fusion of data center network and machine learning, together with some challenges and potential future research opportunities.
\end{abstract}


% \fi
\IEEEpeerreviewmaketitle

\section{Introduction}
\label{Introduction}
\IEEEPARstart{A}
s the storage and computation progressively migrate to the cloud, the data center as the core infrastructure of cloud computing provides vital technical and platform support for enterprise and cloud services.
However, with the rapid rise of the data center scale, the network optimization, resource management, operation and maintenance, and data center security have become more and more complicated and challenging.
What's more, the burgeoning development of 5G has spawned numerous complex, real-time, diversified, and heterogeneous service scenarios \cite{chiaraviglio2018planning,conformance20113rd}, such as enhanced mobile broadband (eMBB) (e.g., ultra-high definition adaptation, augmented reality, virtual reality), ultra reliable low latency communication (uRLLC) (e.g., internet of vehicles, industrial automation, mission-critical applications), and enhanced machine-type communication (eMTC) (e.g., Internet of Things, smart grid, smart cities). 
The emergence of these new services poses new standards and higher requirements for data centers \cite{navarro2020survey,series2015imt}, such as high concurrency, low latency, and micro-burst tolerance.

In terms of DCN automation, benefiting from software defined networks (SDNs), data centers have initially achieved automation in some areas, such as automated installation of network policies and automated network monitoring. However, the implementation of such automation typically depends on predefined policies. Whenever the predefined policies are exceeded, the system lacks adaptive processing capability through autonomous learning, and human intervention must be involved. In the face of these challenges and issues, the traditional solutions \cite{vamanan2012deadline,liu2019task,alizadeh2013pfabric,chiu2017coflourish,yao2017deadline,shieh2011sharing,wang2014rethinking} have become inefficient and incompetent. Moreover, with data availability and security at stake, the issues of data centers are more critical and challenging than ever before \cite{zhang2020rethinking,zhang2018designing}. Driven by these factors, about the last decade both academia and industry have conducted extensive research in improving the intelligence level of DCNs by leveraging machine learning (ML) techniques \cite{lei2019gcn,bezerra2020performance,saber2020online,scherer2015practise,iqbal2020adaptive,ruffy2018iroko}.

% Conceptual_Map_of_Survey
\begin{figure*}
    \centering
    \includegraphics[width=\linewidth]{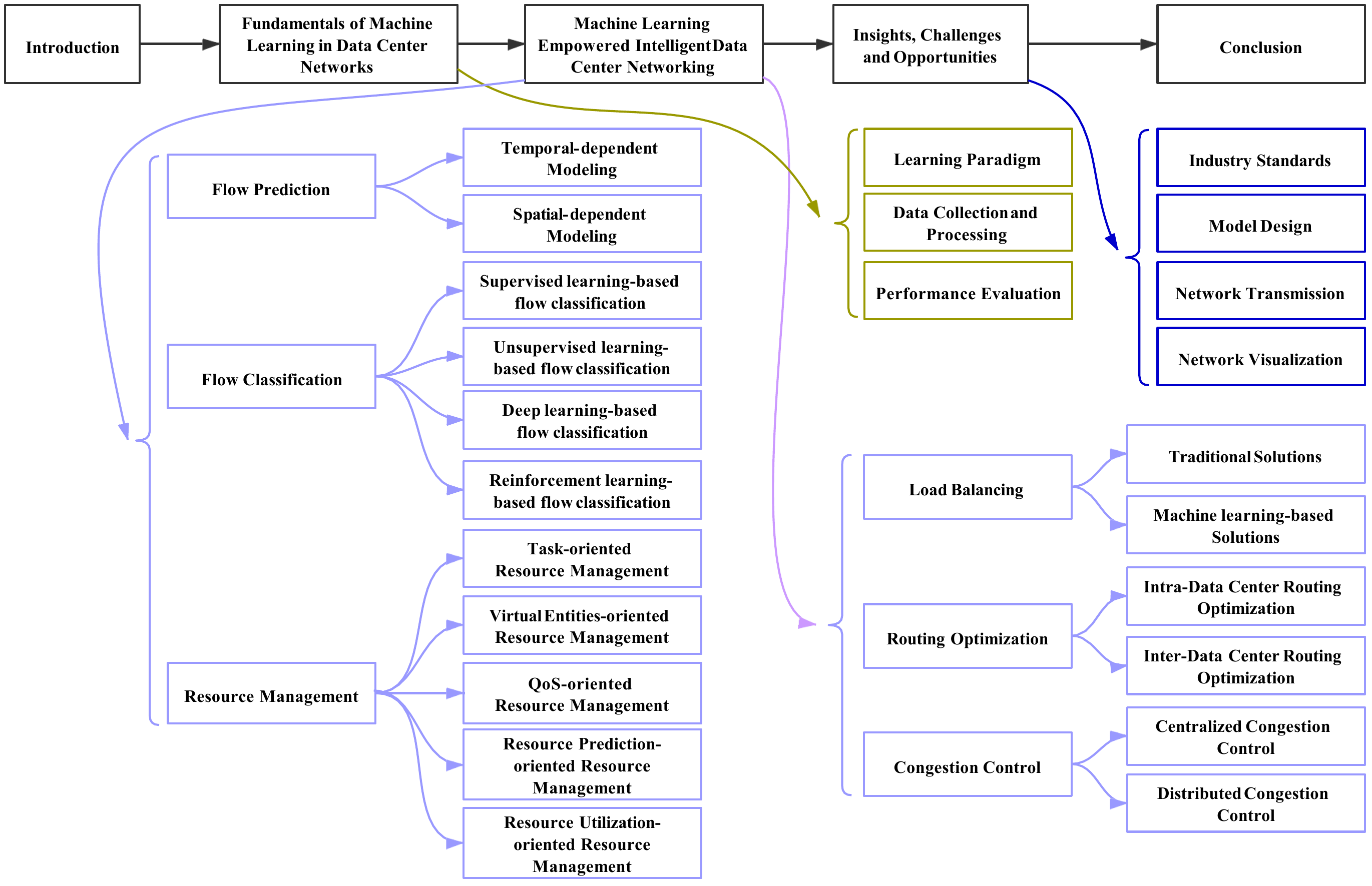}
    \caption{The Conceptual Map of the Survey}
    \label{Conceptual_Map_of_Survey}
    \vspace{-0.3cm}
\end{figure*}

It is universally acknowledged that data-driven ML technologies have made tremendous progress around the last decade \cite{7932863,xie2018survey,8762138}. A quantity of academic research has primarily demonstrated that ML could make more effective decisions and optimizations in the ever-changing network environment. With ML technologies, the vast amount of data accumulated in the network can be well exploited to assist the system in dealing with the complex network problems. The substantial increase in computer storage and processing power (e.g., graphics processing unit (GPU) \cite{google_gpu}, tensor processing unit (TPU) \cite{cloud}) also provides a strong guarantee for ML implementation in DCNs. The introduction of ML technology will greatly help the data center network to improve the network service quality and the efficiency of operation and maintenance (O\&M), so as to cope with the new challenges brought by the increasingly complex network management and dynamic flexible services. With regards to this, various standardization organizations, industries, open-source organizations, and equipment vendors have begun to invest and practice in ML-assisted intelligent data center networking. International standardization organizations such as CCSA and ETSI have started relevant research projects \cite{yang2019federated,dahmen-lhuissier_etsi_nodate}. Open-source organizations such as the Linux Foundation have released several network intelligence related open-source projects. The major operators and equipment vendors have increased their investments and research efforts in network intelligence, and put forward a series of new intelligent networking concepts, such as Juniper's Self-Driving Network \cite{selfdriving}, Gartner's Intent-Based Network System (IBNS) \cite{lerner2017innovation}, Cisco's Intent-Based Network (IBN) \cite{2017intentbased} and Huawei's Intent-Driven Network (IDN) \cite{intentdriven}.

Although the research on intelligent data center networking has made great progress, it is still confronted with many challenges. On the one hand, the strategy of data collection and processing plays an important role in the effectiveness of data-driven ML-based models. In particular, the way of data collection, the impact of the traffic \& computation overhead caused by data collection, and the potential for data leakage are essentially critical. On the other hand, the research on intelligent data center networking is still in the initial stage, where limited by various factors and constraints, the intelligent solutions in some fields are not mature yet, and some intelligence processes are not complete as well. 
For example, flow prediction plays a crucial role in DCN optimization, which servers as a priori knowledge in routing optimization, resource allocation and congestion control. It can grasp the characteristics and trends of network flow in advance, providing necessary support for relevant service optimization and decision-making. Nevertheless, the huge scale of network and the diversity of services impose great challenges in dealing with such flows with irregular and random distributions in both time and space dimensions. 
Flow classification, like flow prediction, is widely used as a priori knowledge for a variety of optimization modules, including flow scheduling, load balancing, and energy management. Regarding the quality of service (QoS), dynamic access control, and resource intelligent optimization, accurate categorization of service flows is critical. According to our research (as shown in Section \ref{Machine Learning Empowered Intelligent Data Center Networking}), the current ML-based traffic classification schemes also have much room for improvement in the fineness of granularity, time efficiency, and robustness. 
Meanwhile, the goal of load balancing is to guarantee a balanced distribution of flows over multiple network routing paths in order to reduce latency, enhance bandwidth usage, and minimize flow completion time. The problem of load balancing is commonly stated as a multi-commodity flow (MCF) problem, which has been proven to be NP-hard. Undoubtedly, the highly dynamic data center (DC) traffic brings great challenges to the load balancing of intra-DC or inter-DC, which requires an efficient grasp of the characteristics of network traffic. 
Simultaneously, resource management, as one of the most important optimization problems in the data center, involves the allocation, scheduling, and optimization of computing, storage, network, and other resources, which has a direct impact on the data center's overall resource utilization efficiency and resource availability, as well as the user experience and revenue of service providers. However, with the increasing complexity of network infrastructure, the explosive growth of the number of hardware devices, and the growing demand for services, the traditional unintelligent solutions can no longer effectively deal with these problems, and there is an urgent need for some intelligent resource management solutions. Homoplastically, routing optimization is also one of the most important research areas and has aroused some discussions in both academia and industry. Routing optimization can benefit from SDN by getting a global view of the network and conveniently deploying techniques, however typical SDN-based solutions cannot sensitively react to real-time traffic changes in data center networks. \cite{amezquita2019efficient,wang2018efficient,xiao2017openflow,guo2018balancing,wang2014freeway,liu2014sdn}. Notably, the resource management and routing optimization should fully consider the diversity of resources and service requirements, whose multi-objective optimization is usually an intractable problem. 
Last but not least, the congestion control mechanism of the network also needs further research in terms of model stability, convergence speed, and robustness. The complexity and diversity of service scenarios and finer granularity of flow demands have made congestion control more complicated in data centers. For instance, some applications require high micro-burst tolerance \cite{shan2017improving,shan2018micro}, while some applications demand low latency \cite{mittal2015timely} or high throughput \cite{gao2017demepro}. Besides, the diverse applications and computing frameworks with different characteristics in data centers further produce a variety of traffic patterns, such as one-to-one, one-to-many, many-to-one, many-to-many, and all-to-all traffic patterns. However, the traditional transmission control protocol-based (TCP-based) solutions are struggle to match all of these diverse traffic patterns' requirements at the same time \cite{flach2013reducing,dong2018pcc}, resulting in queuing delays, jitter incast, throughput collapse, longer flow completion times, and packet loss \cite{choudhury1998dynamic,lu2018dynamic,majidi2019adaptive}.
Above all, the high networking complexity, highly dynamic environment, diverse traffic patterns and diversified services all may make it not so easy to directly employ ML techniques to data center. Here we summarize four key challenges encountered in current research, as follows.

\begin{itemize}
    \item \textbf{Data processing:} The ability of data processing and feature engineering directly impacts the performance of ML algorithms. However, the massive volume of real-time data generated in data centers poses a significant challenge to data processing.
    \item \textbf{ML model selection:} The optimization tasks in data centers are complex and diverse, whereas there is no one-size-fits-all ML model than can efficiently deal with all scenarios. Therefore, how to choose the appropriate ML algorithm for different scenarios and different optimization tasks is a necessary but challenging thing.
    \item \textbf{Collaborative optimization:} Currently, the existing intelligent data center networking solutions usually follow the principle of "one model for one problem". However, the optimization tasks in data centers are numerous with different objectives. Thus, for the scenario of multi optimization tasks, how to achieve an efficient collaborative optimization among multiple intelligent models is a challenging problem.
    \item \textbf{Standardization:} The industry and academia are eagerly waiting for a universally applicable implementation standard to promote DCN intelligence, as many intelligent standards, such as Knowledge-Defined Network (KDN) \cite{mestres2017knowledge}, have not yet been prototyped.
\end{itemize}

In this survey, we comprehensively investigate the research progress of ML-based intelligent data center networking. Figure \ref{Conceptual_Map_of_Survey} shows the organization of this paper. We classify the existing research work into six different fields, namely, flow prediction, flow classification, load balancing, resource management, routing optimization, and congestion control. These existing intelligent DCN solutions in each network area will be analyzed and compared from different dimensions. Furthermore, in-depth insights into the current challenges and future opportunities of ML-assisted DCN will be provided subsequently. The main contributions of this paper are summarized as follows:

\begin{itemize}
    \item [1)]
          To the best of the authors' knowledge, this is the first comprehensive survey about the application of ML in DCNs. We review the peer-reviewed literature published in recent decade, which are of great influence and well received by peers. The diversity of techniques of machine learning is fully respected to ensure a strong support for the subsequent fair comparisons.
    \item [2)]
          We provide enlightening discussions on the usage of ML algorithms in DCNs. We analyze the effectiveness of ML technologies in DCNs from different aspects.
          In order to provide a multi-dimensional and multi-perspective comparison of various solutions, we innovatively propose our REBEL-3S quality assessment criteria.
    \item [3)]
          We identify a number of research challenges, directions and opportunities corresponding to the open or partially solved problems in the current literature.
\end{itemize}

The rest of this paper is organized as follows. First, we briefly introduce some background knowledge about ML and DCNs in Section \ref{Fundamentals of Machine Learning in Data Center Networks}. Then we discuss the wide range of applications of ML in DCNs and provide a comparative analysis from different aspects in Section \ref{Machine Learning Empowered Intelligent Data Center Networking}. In Section \ref{Insights, Challenges and Opportunities}, we provide insights into DCN's intelligence accompanying by some challenges as well as opportunities. Finally, the paper concludes in Section \ref{Conclusion}. For ease of retrieval, the list of abbreviations used in this article is summarized in Table \ref{A List of Key Acronyms}.

\begin{table*}[thp]
    \centering
    \footnotesize
    \caption{A List of Key Acronyms}
    \label{A List of Key Acronyms}
    \begin{tabular}{rlrl}
        \toprule
        AIMD  & Additive Increase Multiplicative Decrease      & LSTM  & Long Short-Term Memory                   \\
        ANN  & Artificial Neural Network            & MA    & Moving Average                      \\
        AR   & Augmented Reality                    & MAE   & Mean Absolute Error                 \\
        ARIMA & Autoregressive Integrated Moving Average Model & MAPE  & Mean Absolute Percentage Error           \\
        ARMA & Autoregressive Moving Average Model  & ME    & Mean Error                          \\
        BGP  & Border Gateway Protocol              & ML    & Machine Learning                    \\
        BI   & Blocking Island                      & MSE   & Mean Squared Error                  \\
        BMP  & BGP Monitoring Protocol              & NBD   & Naïve Bayes discretization          \\
        Bof  & Bag of Flow                          & NFV   & Network Functions Virtualization    \\
        CC   & Congestion Control                   & NIDS  & Network Intrusion Detection System  \\
        CFD  & Computational Fluid Dynamics         & NMSE  & Normalized Mean Square Error        \\
        CNN  & Convolutional Neural Networks        & NN    & Neural Network                      \\
        CRAC & Computer Room Air Conditioner        & O\&M  & Operations and Maintenance          \\
        CRE  & Cognitive Routing Engine             & PAC   & Packaged Air Conditioner            \\
        DBN  & Deep Belief Network                  & PC    & Personal Computer                   \\
        DC   & Data Center                          & QoS   & Quality of Service                  \\
        DCN  & Data Center Network                  & RAE   & Relative Absolute Error             \\
        DDoS & Distributed Denial of Service Attack & RF    & Random Forest                       \\
        DDPG & Deep Deterministic Policy Gradient   & RFR   & Random Forest Regression            \\
        DL   & Deep Learning                        & RL    & Reinforcement Learning              \\
        DNN  & Deep Neural Network                  & RMSE  & Root Mean Squared Error             \\
        DPI  & Deep Packet Inspection               & RNN   & Recurrent Neural Network            \\
        DQN  & Deep Q-Network                       & RRMSE & Relative Root Mean Squared Error    \\
        DRL  & Deep Reinforcement Learning          & RRSE  & Relative Root Squared Error         \\
        DT   & Decision Tree                        & RSNE  & Ratio of Saved Number of Entries    \\
        ELM  & Extreme Learning Machine             & RTT   & Round-Trip Time                     \\
        eMBB & Enhanced Mobile Broadband            & SDN   & Software Defined Network            \\
        eMDI & Enhanced Media Delivery Index        & SL    & Supervised Learning                 \\
        eMTC & enhanced Machine-Type Communication  & SLA   & Service-Level Agreement             \\
        FCT  & Flow Completion Time                 & SNMP  & Simple Network Management Protocol  \\
        FPGA & Field Programmable Gate Array        & SVM   & Support Vector Machine              \\
        FTR  & Fundamental Theory Research          & SVR   & Support Vector Regression           \\
        GBDT  & Gradient Boosting Decision Tree                & TCP   & Transmission Control Protocol            \\
        GNN  & Graph Neural Network                 & TPU   & Tensor Processing Unit              \\
        GRU  & Gated Recurrent Unit                 & TWAMP & Two-Way Active Measurement Protocol \\
        GWO  & Grey Wolf Optimization               & UL    & Unsupervised Learning               \\
        HSMM  & Hidden Semi-markov Model                       & uRLLC & Ultra Reliable Low Latency Communication \\
        IANA & Internet Assigned Numbers Authority  & VC    & Virtual Container                   \\
        IDS  & Intrusion Detection System           & VM    & Virtual Machine                     \\
        IT   & Information Technology               & VN    & Virtual Network                     \\
        AL   & Automatic Learning                   & VR    & Virtual Reality \\
        % \midrule
        CKNN  & \multicolumn{3}{l}{K-nearest Neighbors Traffic Classification With Correlation Analysis}                                                    \\
        \bottomrule
    \end{tabular}
\end{table*}

\section{Fundamentals of Machine Learning in Data Center Networks}
\label{Fundamentals of Machine Learning in Data Center Networks}
%The concept of ML was first introduced in 1959, and then it remained in relative obscurity for a long period. In recent years, ML has attracted the attention of academia and industry again and has been widely studied and employed in various scenarios, ranging from big data analysis and data mining to face recognition. 
In this section, we will briefly review the common learning paradigms of ML and some preliminary knowledge about data collection and processing. Furthermore, to better assess the strengths and weaknesses of the existing research work, we design a multi-dimensional and multi-perspective quality assessment criteria, called REBEL-3S.

\subsection{Learning Paradigm}
\label{Learning Paradigm}
Machine learning paradigms can be generally classified into three categories: supervised learning, unsupervised learning, and reinforcement learning. With the in-depth research and development of ML, some new learning paradigms such as deep learning and deep reinforcement learning have been derived for more complex scenarios.

Supervised learning is a simple and efficient learning paradigm, but it requires data to be labeled, where the manual labeling task is usually complex and time-consuming with a considerable workload. This learning paradigm is mainly used to finish simple tasks, e.g., fault prediction and flow classification. The typical representative supervised models are random forests, SVM, KNN, and decision trees. Unsupervised learning can mine potential hidden structures in unlabeled datasets but are relatively fragile and sensitive to data quality. It is more prone to the interference of anomalous data, and the final learning effect is difficult to quantify. Therefore, few research works applied unsupervised learning in DCNs.

Compared with the former two learning paradigms, deep learning is differentiated and characterized by the depth of learning, where it is expected to find the intrinsic association among data through continuous iterative feature extraction, convolution, pooling and other necessary operations. CNN, RNN, LSTM, and GRU are the most common deep learning models used in data center networks.
Although accompanied by long training time and slow convergence rate which limit its applicability in some scenarios with high real-time requirements, they are widely used owing to their excellent performance \cite{cao2017interactive,zhu2020differentiated,liu2020fine,lu2020ai}.

Unlike the previous models, reinforcement learning is designed to self-learn through continuous interactions with the external environment. The actions performed by the agent can adjust themselves according to the feedback (reward or punishment) given by the environment so as to achieve the global optimal effect. In view of the strong adaptive self-learning ability of reinforcement learning,  it has been broadly applied to solve complex problems such as congestion control, routing optimization, and flow scheduling. However, reinforcement learning also has its own problems:
(1) The learning model tends to fall into local optimal solutions.
(2) The learning model usually requires a long training time, challenging to meet the need of real-time requirements.
(3) The learning results may have overfitting phenomenon, which will result in poor model generalization ability in the face of new complex environment.

Deep learning has strong perception ability, but it lacks certain decision-making ability, while reinforcement learning has decision-making ability, yet it has nothing to do with perception problems. Therefore, deep reinforcement learning is proposed by combining the complementary advantages of two learning paradigms, to enable the model with both perceptual ability and decision-making ability. Unfortunately, deep reinforcement learning still retains the problems of poor stability and complex reward functions.

It can be seen that different learning paradigms have specific limitations, and how to adaptively select appropriate ML models for complex optimization tasks with different objectives in data centers is extremely vital and is much difficult.

\subsection{Data Collection And Processing}
Data collection and processing is regarded as the first step to realize the intelligence of ML-assisted data centers, while the quality of source data directly determines the performance of ML models. However, in the complex and dynamic data center network environment, the massive data generated in real time are usually transient, multidimensional, heterogeneous and diversified, which brings great challenges to the data collection and processing. In this section, we will present our investigation findings and analysis of data collection and processing in data centers from three aspects, namely, data collection scenarios, data collection techniques, and feature engineering. Finally, we will provide our insights into the open problems and challenges in this field.

\subsubsection{Data Collection Scenarios}
\label{Data collection scenarios}
Data collection scenarios in DCNs can be divided into four categories: service data collection, protocol data collection, network performance data collection, and basic network data collection.
The service data includes the information of service SLA, and service topology. The service SLA can be further divided into flow-level SLA and service-level SLA. Flow-level SLA measurement is mainly conducted through IFIT (In-situ Flow Information Telemetry) and eMDI. Service-level SLA data can be collected through TWAMP/Y.1731.
The protocol data includes protocol stack states, routing information, and delay statistics.
The network performance data collection typically consists of interface statistics, queue statistics, and network element health data.
The basic network data collection mainly gathers the information of the physical topology, alarms, and logs.

\subsubsection{Data Collection Techniques}
\label{Data Collection Technology}
In the dynamic data center environment, the decision-making of network optimization strategies has strict requirements on the timeliness and quality of the collected data, which also poses a great challenge to the data collection techniques.
Empirically, different scenarios have different quality requirements of data collection, thus different data collection techniques will be adopted, accordingly. Generally, the data collection techniques can be grouped into three categories: real time data collection, protocol data collection, and basic data collection.
The real-time data collection techniques mainly include Telemetry, IFA, Netflow, sFlow and OpenFlow, which collect the data with a time granularity of seconds reflecting the real time network status.
The protocol data collection techniques target at collecting the routing protocol (e.g., BGP) data as well as the topology information, where the typical representatives are BGP-LS and BMP.
The SNMP and Syslog are usually employed as the basic data collection techniques to provide basic network information, such as network log, and alarm.

\subsubsection{Feature Engineering}
\label{Feature Engineering}
Feature engineering refers to the process of transforming the original raw data into the training data, and it directly determines the effectiveness of one ML model, where it can reduce the dimension of data lowering the computing cost. Its essential goal is to improve the performance of the model by acquiring better features of the training data.
The feature engineering actions primarily include feature selection and feature extraction. Comparatively, the feature selection plays a more vital role, where high-quality feature selection is helpful to remove redundant irrelevant features and improve the accuracy of ML models. Similar to the backbone internet scenario \cite{fraleigh2003packet,barakat2003modeling,tao2003methodology}, in DCN scenario, feature engineering can be divided into three different granularity levels: packet-level, flow-level, and application stream-level.

The most fine-grained packet-level features collect packet information and statistics. The flow-level features are generally represented as a 5-tuple, i.e. <source IP address, source port number, destination IP address, destination port number, transport layer protocol>, and flows at this level are usually classified according to the transport layer protocol. The application stream-level is characterized by the number of flows in the Bag of Flow (BoF) level \cite{zhang2012classification,zhang2012internet}, which can be represented as a 3-tuple, i.e., <source IP address, destination IP address, transport layer protocol>. It is suitable for studying the long-term flow statistics of the backbone network at a coarser granularity, but the collection of such high-quality data can increase the computational overhead of the data center.

Feature selection should be adapted to service scenarios. However, in data centers, feature selection strategies are usually not deterministic or invariant. Even for the same problem in the same scenario, features may be inconsistent across different solutions. For example, on the issue of reducing the energy consumption of the data center, Sun et al. \cite{sun2020smartfct} simply took the temperature of the chassis as the feature input, while Yi et al. \cite{yi2020efficient} considered the interaction between the average utilization, temperature, and energy consumption of the processor, further expanding the number of key features and effectively exploring the relationship between the features. In conclusion, there are various strategies for feature selection, and the differentiated features will have a direct impact on the final results of the model.

\subsubsection{Challenges and insights}
\label{Challenges and insights}
Through the above investigation and analysis, we have summarized the following challenges in current data collection and processing.
\begin{itemize}
    \item \textbf{Data collection load:} The volume of data in DCN has been growing explosively in recent years \cite{synergy_research_group}, which already reached 403 exabytes in 2021 \cite{noauthor_big_nodate}. Such massive data bring many problems to data acquisition. For example, whether collecting large volume of data will cause congestion within the collection devices, and whether the data with critical features can be accurately collected. When the data are distributed on different paths, the same data will be collected multiple times on different links if collection points are deployed on all links. Thus, effectively reducing the duplicate sampling of data can help relieve the burden and economic overhead of data collection, and data collection should minimize the impact on the network.
    \item \textbf{Data collection methods:} The dynamic, diversified, complex and real-time DCN environment imposes great challenges to the data collection. Although there are many kinds of data collection methods, they cannot be always well aligned with the collection needs. To put it in practical terms, due to technical implementation limitations, Netstream can only analyze the basic 5-tuple information, and changes beyond the IP header cannot be collected and analyzed. At the same time, the visual display of network data collection results deserve more profound research.
    \item \textbf{Data security and privacy:} The security and privacy of the collected data have become the primary focus of data center networks. A series of behaviors such as data desensitization, access control, and leakage prevention are major issues in the current data center networks.
\end{itemize}

% The_Confusion_Matrix_For_Binary_Classification
%\begin{figure}
%    \centering
%    \includegraphics[width=0.6\linewidth]{./The_Confusion_Matrix_For_Binary_Classification.pdf}
%    \caption{The Confusion Matrix For Binary Classification}
%    \label{The_Confusion_Matrix_For_Binary_Classification}
%    \vspace{-0.3cm}
%\end{figure}

\subsection{Performance Evaluation of ML-based Solutions In DCN}
\label{Performance Evaluation}

\begin{table}[H]
    \centering

    \caption{The Meaning of The REBEL-3S}
    \label{The Meaning of The REBEL-3S}
    \begin{tabular}{c | c}
        \hline
        \textbf{Abbreviations} & \textbf{Properties}              \\
        \hline
        \textbf{R}                      & \textbf{R}eliability                      \\
        \hline
        \textbf{E}                      & \textbf{E}nergy Efficiency                \\
        \hline
        \textbf{B}                      & \textbf{B}andwidth Utilization            \\
        \hline
        \textbf{L}                      & \textbf{L}atency                          \\
        \hline
        \textbf{3S}                     & \textbf{S}ecurity, \textbf{S}tability, \textbf{S}calability \\
        \hline
    \end{tabular}
\end{table}

The performance evaluation of a model should fully consider its specific application scenarios. To provide a multi-dimensional and multi-perspective comparison of various intelligent solutions in the DCN scenario, we propose a quality assessment criteria named REBEL-3S, as illustrated in Table \ref{The Meaning of The REBEL-3S}.
"R" stands for reliability, which refers to the robustness and availability of a solution, including the capabilities of failure detection, fault tolerance, and self-healing, etc.
"E" stands for energy efficiency, which refers to whether the solution considers the energy cost.
"B" and "L" represent bandwidth utilization and latency, respectively.
"3S" means security, stability, and scalability of the network, i.e., whether to consider security and privacy against anonymous attacks and abnormal flows, whether to consider network fluctuations and seasonal variation of flows, and whether to consider scalability performance of the solution. It will be marked "YES" to indicate that the solution has taken the above evaluation dimensions into account and vice versa with "NO".

\section{Machine Learning Empowered Intelligent Data Center Networking}
\label{Machine Learning Empowered Intelligent Data Center Networking}
Machine learning has been widely studied and practiced in data center networks, and a large number of achievements have been made. In this section, we will review, compare, and discuss the existing work in the following research areas: flow prediction, flow classification, load balancing, resource management, energy management, routing optimization, and congestion control.

\subsection{Flow Prediction}
\label{Flow Prediction}
Flow prediction plays a crucial role in DCN optimization, and servers as a priori knowledge in routing optimization, resource allocation and congestion control. It can grasp the characteristics and trends of network flow in advance, providing necessary support for relevant service optimization and decision-making. However, the huge scale of network and the diversity of services impose great challenges in dealing with such flows with irregular and random distributions in both time and space dimensions. For instance, the flow estimation methods based on the flow gravity model \cite{zhang2009spatio,zhang2003fast} and network cascade imaging \cite{headquarters2007cisco,qiao2013efficient} are challenging to cope with a large number of redundant paths among the massive number of servers in data centers.

The current research work can be generally divided into classical statistical models and ML-based prediction models. The classical statistical models usually include autoregressive models (AR), moving average models (MA), autoregressive moving average models (ARMA) \cite{anderson1971statistical}, and autoregressive synthetic moving average models (ARIMA) \cite{feng2005study}. These models cannot cope with high-dimensional and complex nonlinear relationships yet, and their efficiency and performance in complex spaces are fairly limited. ML-based prediction models can be trained based on historical flow data information to find potential logical relationships in complex and massive data, explaining the irregular distribution of network flow in time and space. According to the flow's spatial and temporal distribution characteristics, we classify ML-based prediction solutions into temporal-dependent modeling and spatial-dependent modeling. Next, we will discuss and compare the existing representative work of these two schemes from different perspectives, followed by our insights into flow prediction.

\subsubsection{Temporal-dependent Modeling}
The temporal-dependent modeling focuses on the temporal dimension inside the data center. Flow forecasting is usually achieved by using one-dimensional time series data. Liu et al. \cite{liu2017adaptive} proposed an elephant flow detection mechanism. They first predicted future flow based on dynamical traffic learning (DTL) algorithm and then dynamically adjusted the elephant flow judgment threshold to improve detection accuracy. However, the frequent involvement of the controller causes extra computational and communication overhead. Beyond these, researchers have also made great efforts to optimize the prediction accuracy with a finer granularity. Szostak et al. \cite{szostak2020application} used supervised learning and deep learning algorithms to predict future flow in dynamic optical networks. They tested six ML classifiers based on three different datasets. Hardegen et al. \cite{hardegen2020predicting} collected about 100,000 flow data from a university DCN and used deep learning to perform a more fine-grained predictive analysis of the flow. Besides, researchers \cite{wang2017effieye,mbous2019kalman,liu2019sdn,aibin2018traffic} have also carried on a lot of innovative work on the basic theoretical research of artificial intelligence. However, some of the experiments to verify the effectiveness of these intelligent schemes are not sufficient. For example,  Hongsuk et al. \cite{yi2017deep} only conducted experimental comparisons on the effectiveness with different parameter settings, lacking the cross-sectional comparisons as aforementioned.

\subsubsection{Spatial-dependent Modeling}
The spatial-dependent modeling focuses on both temporal and spatial dimensions across data centers. Nearly 67\% of these intelligent solutions compare with the classical statistical models and other ML-based prediction models. The spatial-dependent modeling greatly improves the feasibility and accuracy of solutions, but it also increases the complexity and the operational cost of network O\&M. According to our investigations, Over 45\% of commercial data center traffic prediction schemes adopt spatial-dependent modeling. Li et al. \cite{li2016predicting} studied flow transmission schemes across data centers, combined wavelet transform technique with a neural network, and used the interpolation filling method to alleviate the monitoring overhead caused by the uneven spatial distribution of data center traffic. Its experiments conducted in Baidu data center showed that the scheme could reduce the prediction error by 5\% to 30\%. We also note that the about 70\% of intelligent flow prediction solutions used real-world data.
Pfülb et al. \cite{pfulb2019study} based on the real-world data obtained from a university data center that had been desensitized and visualized, the authors used deep learning to predict the inter-DC traffic.

\subsubsection{Discussion and Insights}
The comprehensive comparisons of the existing approaches are detailed in Tables \ref{flow_prediction_1} and \ref{flow_prediction_2}. Some of our insights into flow prediction are as below.

\begin{itemize}
    \item \textbf{Prior knowledge.} ML algorithms such as Support Vector Regression (SVR) \cite{chen2015forecasting} and Random Forest Regression (RFR) \cite{johansson2014regression}, compared to classical statistical models, can handle high-dimentional data and obtain their nonlinear relationships well. Nevertheless, their performance in exceptionally complex spatio-temporal scenarios is still limited, partially because they require additional expert knowledge support, where the model learns through the features pre-designed by the experts. However, these features usually can not fully describe the data's essential properties.
    \item \textbf{Quality of source data.} The performance of flow prediction heavily depends on the quality of source data, with respect to authenticity, validity, diversity, and instantaneity. Not only for flow prediction, the quality of source data also plays a crucial role in other optimization scenarios of intelligent data center network, which we will explain in section \ref{Data Quality Assessment Standards}.
    \item \textbf{Anti-interference ability.} The network upgrading, transformation and failures typically can cause sudden fluctuation of traffic, and these abnormal data will interfere with the ultimate accuracy of the model. In order to improve the accuracy of traffic prediction, it is suggested to provide an abnormal traffic identification mechanism to identify the abnormal interference data and eliminate them when executing traffic predictions.
\end{itemize}

% flow prediction
\begin{sidewaystable*}[thp]
    \caption{Research Progress of Data Center Network Intelligence: Flow Prediction} % title name of the table
    \label{flow_prediction_1}
    \centering % centering table
    \begin{threeparttable}
        \resizebox{\linewidth}{!}{
            \begin{tabular}{p{3.5cm} p{1.5cm}<{\centering} p{2.5cm} p{7cm} p{2cm} p{7cm} p{3cm}<{\centering}}
                \hline\hline % inserting double-line
                \specialrule{0em}{1pt}{1pt}
                \textbf{Ref}                                                    & \textbf{Category \tnote{1}}                                                                                                                                                                           & \textbf{ML Category \& Basic Model} & \textbf{Features} & \textbf{Estimation Function} & \textbf{Data Source} & \textbf{Experimental Comparison Subjects \tnote{2}} \\
                \specialrule{0em}{1pt}{1pt}
                \hline % inserts single-line
                \specialrule{0em}{1pt}{1pt}
                % Entering 1st row
                % Çetiner et al. \cite{ccetiner2010neural}
                %                                                                 & S
                %                                                                 & DL, DNN
                %                                                                 & Prediction of the flow volume based on the historical data in each major junction in the city
                %                                                                 & Accuracy
                %                                                                 & Over 130,000 real historical data provided by the semi-governmental organization ISBAK
                %                                                                 & \CheckmarkBold  |  \XSolidBrush |  \XSolidBrush                                                                                                                                                                                                                                                                                                                                                                                                                                                                                                                                                                                                                                    \\
                \specialrule{0em}{1pt}{1pt}
                \hline % inserts single-line
                \specialrule{0em}{1pt}{1pt}
                % Entering 2st row
                Pfülb et al. \cite{pfulb2019study}
                                                                                & S
                                                                                & DL, DNN
                                                                                & Trained with a large dataset of approximately 50 million streams for more granular traffic prediction and visual analysis
                                                                                & Accuracy, etc.
                                                                                & Collected data from the networks at Fulda University of Applied Sciences
                                                                                & \CheckmarkBold  |  \XSolidBrush  |  \XSolidBrush                                                                                                                                                                                                                                                                                                                            \\
                \specialrule{0em}{1pt}{1pt}
                \hline % inserts single-line
                \specialrule{0em}{1pt}{1pt}
                % Entering 3st row
                Liu et al. \cite{liu2017adaptive}
                                                                                & T
                                                                                & FTR, DTL
                                                                                & Adopted dynamical flow learning (DTL) algorithm, weighted optimization based on Gaussian distribution and smooth mechanism based on difference estimation
                                                                                & Customized
                                                                                & CAIDA \cite{analysiscaida}
                                                                                & \CheckmarkBold  |  \XSolidBrush  |  \CheckmarkBold                                                                                                                                                                                                                                                                                                                          \\

                \specialrule{0em}{1pt}{1pt}
                \hline % inserts single-line
                \specialrule{0em}{1pt}{1pt}
                % Entering 4st row
                Szostak et al. \cite{szostak2019machine,szostak2020application} & T
                                                                                & SL/DL, KNN etc.
                                                                                & A traffic prediction method in dynamic optical networks for serving VNF was proposed
                                                                                & Customized
                                                                                & Simulated data
                                                                                & \CheckmarkBold  |  \XSolidBrush  |  \CheckmarkBold                                                                                                                                                                                                                                                                                                                          \\
                \specialrule{0em}{1pt}{1pt}
                \hline % inserts single-line
                \specialrule{0em}{1pt}{1pt}
                % Entering 5st row
                Yi et al. \cite{yi2017deep}                                     & T
                                                                                & DL, DNN
                                                                                & The first research team to use TensorFlow-based DNN for traffic prediction
                                                                                & Customized
                                                                                & Obtained from about 0.5 million probe
                vehicles with an on-board device (OBD)
                                                                                & \XSolidBrush  |  \XSolidBrush  |  \CheckmarkBold                                                                                                                                                                                                                                                                                                                            \\
                \specialrule{0em}{1pt}{1pt}
                \hline % inserts single-line
                \specialrule{0em}{1pt}{1pt}
                % Entering 6st row
                Wang et al. \cite{wang2017effieye}                              & T
                                                                                & FTR, None
                                                                                & Installed preclassified information into the controller for fast classification and use OpenFlow for event management
                                                                                & None
                                                                                & No experiments
                                                                                & \XSolidBrush  |  \XSolidBrush  |  \XSolidBrush                                                                                                                                                                                                                                                                                                                              \\
                \specialrule{0em}{1pt}{1pt}
                \hline % inserts single-line
                \specialrule{0em}{1pt}{1pt}
                % Entering 7st row
                Hardegen et al. \cite{hardegen2019flow}
                                                                                & S
                                                                                & DL, DNN
                                                                                & Unlike the binary classification of "mice" and "elephant", the authors triple-classified according to the predicted bit rate, while using pre-processing, anonymization, and visualization techniques
                                                                                & Accuracy, etc.
                                                                                & Data collection took place in a real-world production network at Fulda University of Applied Sciences
                                                                                & \CheckmarkBold  |  \XSolidBrush  |  \XSolidBrush                                                                                                                                                                                                                                                                                                                            \\
                \specialrule{0em}{1pt}{1pt}
                \hline % inserts single-line
                \specialrule{0em}{1pt}{1pt}
                % Entering 8st row
                Mozo et al. \cite{mozo2018forecasting}
                                                                                & S
                                                                                & DL, CNN
                                                                                & Multiresolution strategy multiple-channel convolutions to incorporate multiresolution context
                                                                                & MSE/MAE
                                                                                & ONTS \cite{noauthor_online_nodate}
                                                                                & \CheckmarkBold  |  \CheckmarkBold  |  \CheckmarkBold                                                                                                                                                                                                                                                                                                                        \\
                \specialrule{0em}{1pt}{1pt}
                \hline % inserts single-line
                \specialrule{0em}{1pt}{1pt}
                % Entering 9st row
                Lei et al. \cite{lei2019gcn}
                                                                                & S
                                                                                & DL, LSTM, etc.
                                                                                & The dynamic prediction of the network was formulated as a temporal link prediction, combined various NN structures, took various evaluation measures, and conducted adequate experiments
                                                                                & Customized
                                                                                & UCSB\cite{ramachandran2007routing}, KAIST\cite{lee2009slaw}, BJ-Taxi\cite{yuan2011driving}, and NumFabric\cite{nagaraj2016numfabric}
                                                                                & \CheckmarkBold  |  \CheckmarkBold  |  \CheckmarkBold                                                                                                                                                                                                                                                                                                                        \\
                \specialrule{0em}{1pt}{1pt}
                \hline % inserts single-line
                \specialrule{0em}{1pt}{1pt}
                % Entering 10st row
                Cao et al. \cite{cao2017interactive}
                                                                                & S
                                                                                & DL, GRU/CNN
                                                                                & Learned network flow as images to capture the network-wide services’ correlations
                                                                                & RMSE
                                                                                & Historical data from Yahoo's Data Center
                                                                                & \CheckmarkBold  |  \CheckmarkBold  |  \CheckmarkBold
                \\
                \specialrule{0em}{1pt}{1pt}
                \hline % inserts single-line
                \specialrule{0em}{1pt}{1pt}
                % Entering 11st row
                Li et al. \cite{li2016inter,li2016predicting}
                                                                                & S
                                                                                & DL, DNN
                                                                                & Combine wavelet transform with DNN to improve prediction accuracy
                                                                                & RRMSE
                                                                                & A production data center with tens of thousands of servers from Baidu for six weeks
                                                                                & \CheckmarkBold  |  \CheckmarkBold   |  \CheckmarkBold                                                                                                                                                                                                                                                                                                                       \\
                \specialrule{0em}{1pt}{1pt}
                \hline % inserts single-line
                \specialrule{0em}{1pt}{1pt}
                % Entering 12st row
                Mbous et al. \cite{mbous2019kalman}
                                                                                & T
                                                                                & FTR, None
                                                                                & Kalman filter-based algorithm
                                                                                & Utilization rate, etc.
                                                                                & Simulated data
                                                                                & \CheckmarkBold   |  \CheckmarkBold   |  \XSolidBrush
                \\
                \specialrule{0em}{1pt}{1pt}
                \hline % inserts single-line
                \specialrule{0em}{1pt}{1pt}
                % Entering 13st row
                Shi et al. \cite{shi2018lstm}
                                                                                & T
                                                                                & DL, LSTM
                                                                                & Traffic forecasting in hybrid data center networks to aid with optical path reconfiguration
                                                                                & MSE
                                                                                & Simulated data
                                                                                & \CheckmarkBold  |  \XSolidBrush  |  \XSolidBrush                                                                                                                                                                                                                                                                                                                            \\
                \specialrule{0em}{1pt}{1pt}
                \hline % inserts single-line
                \specialrule{0em}{1pt}{1pt}
                % Entering 14st row
                Balanici et al.  \cite{balanici2019machine,balanici2019multi,balanici2019server}
                                                                                & T
                                                                                & DL, DNN
                                                                                & Primary Application Solution
                                                                                & MSE, MAE
                                                                                & Mix of simulated and real-world data
                                                                                & \CheckmarkBold  |  \XSolidBrush  |  \XSolidBrush                                                                                                                                                                                                                                                                                                                            \\
                \specialrule{0em}{1pt}{1pt}
                \hline\hline % inserting double-line
            \end{tabular}
        }
        \begin{tablenotes}
            \footnotesize
            \item[1] For simplicity, "S" indicates Spatial-dependent Modeling and "T" denotes Temporal-dependent Modeling.
            \item[2] The notations \CheckmarkBold and \XSolidBrush in this column are used to indicate whether the scheme has carried out the specified experiments. As shown in this column, from left to right, it represents self-comparison with\\ different parameters, comparisons with traditional schemes, and comparisons with intelligent schemes, respectively. Unless otherwise specified, it has the same meaning in subsequent tables.
        \end{tablenotes}
    \end{threeparttable}
\end{sidewaystable*}

% flow prediction
\begin{sidewaystable*}[thp]
    \newcommand{\tabincell}[2]{\begin{tabular}{@{}#1@{}}#2\end{tabular}}
    \caption{Research Progress of Data Center Network Intelligence: Flow Prediction} % title name of the table
    \label{flow_prediction_2}
    \centering % centering table
    \begin{threeparttable}
        \resizebox{ \linewidth}{!}{
            \begin{tabular}{p{3.5cm} p{1.5cm}<{\centering} p{2.5cm} p{7cm} p{2cm} p{7cm} p{3cm}<{\centering}}
                \hline\hline % inserting double-line
                \specialrule{0em}{1pt}{1pt}
                \textbf{Ref} & \textbf{Category \tnote{1}}                                                                                                                                                         & \textbf{ML Category \& Basic Model} & \textbf{Features} & \textbf{Estimation Function} & \textbf{Data Source} & \textbf{Experimental Comparison Subjects} \\
                \specialrule{0em}{1pt}{1pt}
                \hline % inserts single-line
                \specialrule{0em}{1pt}{1pt}
                % Entering 15st row
                Singh et al. \cite{singh2018machine}
                             & S
                             & DL, LSTM
                             & A multi-class dynamic service model was considered
                             & RMSE
                             & Simulated data
                             & \CheckmarkBold  |  \XSolidBrush  |  \XSolidBrush                                                                                                                                                                                                                                                                                                \\
                \specialrule{0em}{1pt}{1pt}
                \hline % inserts single-line
                \specialrule{0em}{1pt}{1pt}
                % Entering 16st row
                Estrada-Solano et al. \cite{estrada2019nelly}
                             & S
                             & DL, DT
                             & Incremental learning-based network flow prediction
                             & FPR, etc.
                             & University data centers \cite{benson_data_2010}
                             & \XSolidBrush  |  \CheckmarkBold  |  \CheckmarkBold                                                                                                                                                                                                                                                                                              \\
                \specialrule{0em}{1pt}{1pt}
                \hline % inserts single-line
                \specialrule{0em}{1pt}{1pt}
                % Entering 17st row
                Bezerra et al. \cite{bezerra2020performance}
                             & S
                             & DL, RNN
                             & Proposed a hybrid prediction model based on FARIMA and RNN models
                             & RMSE, MAPE, etc.
                             & Historical data from Facebook's Data Center
                             & \CheckmarkBold  |  \CheckmarkBold  |  \CheckmarkBold                                                                                                                                                                                                                                                                                            \\
                \specialrule{0em}{1pt}{1pt}
                \hline % inserts single-line
                \specialrule{0em}{1pt}{1pt}
                % Entering 18st row
                Hardegen et al. \cite{hardegen2020predicting}
                             & T
                             & DL, DNN
                             & More granular network flow prediction
                             & Accuracy, etc.
                             & University network flow data, about 100,000 records
                             & \CheckmarkBold  |  \XSolidBrush  |  \XSolidBrush                                                                                                                                                                                                                                                                                                \\
                \specialrule{0em}{1pt}{1pt}
                \hline % inserts single-line
                \specialrule{0em}{1pt}{1pt}
                % Entering 19st row
                Yu et al. \cite{yu2019scheduling}
                             & T
                             & DL, RNN/LSTM
                             & Achieved four classification based on time and frequency, combined with RNN and LSTM to propose a new flow prediction method
                             & Customized
                             & Simulated data
                             & \CheckmarkBold   |  \CheckmarkBold  |   \CheckmarkBold                                                                                                                                                                                                                                                                                          \\
                \specialrule{0em}{1pt}{1pt}
                \hline % inserts single-line
                \specialrule{0em}{1pt}{1pt}
                % Entering 20st row
                Liu et al. \cite{liu2019sdn}
                             & T
                             & FTR, Unspecified
                             & Learning multiple historical traffic matrixes (TMs) through gradient boosting machine (GBM) method rather than monitoring each flow
                             & NMSE, RSNE
                             & Real-world data from unknown sources
                             & \CheckmarkBold  |  \CheckmarkBold |  \XSolidBrush                                                                                                                                                                                                                                                                                               \\
                \specialrule{0em}{1pt}{1pt}
                \hline % inserts single-line
                \specialrule{0em}{1pt}{1pt}
                % Entering 21st row
                Luo et al. \cite{luo2019spatiotemporal}
                             & S
                             & SL/DL, KNN/LSTM
                             & A hybrid flow prediction methodology, highlights the importance of the highly relevant stations to the prediction result
                             & RMSE, Accuracy
                             & Provided by the Transportation Research Data Lab (TDRL)
                             & \CheckmarkBold  |  \CheckmarkBold  |  \CheckmarkBold                                                                                                                                                                                                                                                                                            \\
                \specialrule{0em}{1pt}{1pt}
                \hline % inserts single-line
                \specialrule{0em}{1pt}{1pt}
                % Entering 22st row
                Nie et al. \cite{nie2016traffic}
                             & T
                             & DL, DBN
                             & A flow prediction method based on DBN was proposed and learned the statistical properties between flow start and end nodes. A network tomography model was also constructed
                             & Standard deviation, etc.
                             & Real-world network flow data from unknown sources
                             & \XSolidBrush  |  \CheckmarkBold  |  \CheckmarkBold                                                                                                                                                                                                                                                                                              \\
                \specialrule{0em}{1pt}{1pt}
                \hline % inserts single-line
                \specialrule{0em}{1pt}{1pt}
                % Entering 23st row
                Aibin, Michal et al.  \cite{aibin2018traffic,aibin2018traffica}
                             & S
                             & FTR, MCTS
                             & An evaluation of a specific provider-centric use case for control and provisioning of DC services in WANs.
                             & Decision time etc
                             & Simulated data
                             & \CheckmarkBold  |  \CheckmarkBold  |  \CheckmarkBold                                                                                                                                                                                                                                                                                            \\
                \specialrule{0em}{1pt}{1pt}
                \hline % inserts single-line
                \specialrule{0em}{1pt}{1pt}
                % Entering 24st row
                Yu et al. \cite{yu2020traffic}
                             & T
                             & SL/DL, SNN
                             & Proposed a supervised spiking NN (s-SNN) framework with multi-synaptic mechanism and error feedback model for flow prediction in hybrid E/O switching intra-datacenter network
                             & Resource occupation rate, etc.
                             & Packet head information collected every five minutes for 10 days from three university data centers in Beijing, China
                             & \CheckmarkBold  |  \XSolidBrush  |  \CheckmarkBold                                                                                                                                                                                                                                                                                              \\
                \specialrule{0em}{1pt}{1pt}
                \hline % inserts single-line
                \specialrule{0em}{1pt}{1pt}
                % Entering 25st row
                Paul et al. \cite{paul2019traffic}
                             & T
                             & DL, LSTM/GRU
                             & Analysed the performances of different RNN models with activation functions to obtain future flow demands.
                             & RMSE, SMAPE
                             & Used part of the data released by Telecom Italia in 2015
                             & \CheckmarkBold  |   \XSolidBrush   |  \XSolidBrush                                                                                                                                                                                                                                                                                              \\
                \specialrule{0em}{1pt}{1pt}
                \hline % inserts single-line
                \specialrule{0em}{1pt}{1pt}
                % Entering 26st row
                Guo et al. \cite{guo2018deep}
                             & S
                             & DL, DNN
                             & Designed an adaptive and scalable downlink based flow predictor that exploits the temporal and spatial characteristics of inter-DC flows and provides accurate and timely forecasts
                             & SSE
                             & Simulated data
                             & \CheckmarkBold  |  \XSolidBrush   |  \XSolidBrush                                                                                                                                                                                                                                                                                               \\
                \specialrule{0em}{1pt}{1pt}
                \hline % inserts single-line
                \specialrule{0em}{1pt}{1pt}
                % Entering 27st row
                Zhu et al. \cite{zhu2019machine}
                             & T
                             & DL/UL, LSTM/BIRCH
                             & Used LSTM based NN to predict the arrival of jobs and aggregate requests for computing resources
                             & MSE, etc.
                             & ClusterData2011\_2 \cite{reiss2011google}, which is a data set released by Google Data Center
                             & \CheckmarkBold  |  \XSolidBrush  |  \CheckmarkBold                                                                                                                                                                                                                                                                                              \\
                \hline\hline % inserting double-line
            \end{tabular}
        }
        \begin{tablenotes}
            \footnotesize
            \item[1] For convenience, we use "S" for Spatial-dependent Modeling and "T" for Temporal-dependent Modeling.
        \end{tablenotes}
    \end{threeparttable}
\end{sidewaystable*}

\subsection{Flow Classification}
\label{Flow Classification}
Similar to flow prediction, flow classification is also widely used as a priori knowledge for many other optimization modules such as flow scheduling, load balancing, and energy management. Accurate classification of service flows is essential for QoS, dynamic access control, and resource intelligent optimization. The daily operation and maintenance also require accurate classification of unknown or malicious flows. Moreover, a reasonable prioritized classification ordering can help enterprise network operators optimize service applications individually and meet the resource management requirements and service needs. Nevertheless, the highly dynamic and differentiated traffic, and complex traffic transmission mechanism greatly increases the difficulty of traffic classification.

Traditional traffic classification schemes are usually based on the information of port, payload, and host behaviors. In the early stages of the Internet, most protocols used well-known port numbers assigned by the Internet Assigned Numbers Authority (IANA) \cite{service}. However, protocols and applications began to use random or dynamic port numbers so as to hide network security tools. Some experimental results further show that port-based classification methods are not very effective, for example, Moore et al. \cite{moore2005toward} observed that the accuracy of the classification techniques based on IANA port list does not exceed 70\%. To overcome the limitations of the above classification techniques, the payload-based flow classification method was proposed as an alternative. The payload-based approach, also known as deep packet inspection (DPI), classifies flows by examining the packet payload and comparing it with the protocols' known signatures \cite{finsterbusch2013survey,sen2004accurate,goo2016payload,fu2017efficient}. Common DPI tools include L7 filter \cite{application} and OpenDPI \cite{bhatia2021thomasbhatia}. However, such DPI-based solutions incur high computation overhead and storage cost though they can achieve higher accuracy of traffic classification than port-based soutions. Although the accuracy is improved compared to the previous methods, the complexity and computational effort are significantly higher. Furthermore, dealing with the increasingly prominent network privacy and security issues also brings high complexity and difficulty to DPI-based techniques \cite{bernaille2007implementation,erman2007identifying}. Thus, some researchers put forward a new kind of flow classification technique based on host behaviors. This technique uses the hosts' inherent behavioral characteristics to classify flows, overcoming the limitations caused by unregistered or misused port numbers and high loads of encrypted packets. Nevertheless, the location of the monitoring system largely determines the accuracy of this method \cite{boutaba2018comprehensive}, especially when the observed communication patterns may be affected by the asymmetry of routing.

In the face of such dilemma of traditional solutions, ML-based flow classification techniques can address the mentioned limitations effectively \cite{dainotti2012issues,xue2013traffic}. Based on the statistical characteristics of data flows, they complete the complex classification tasks with a lower computational cost. Next, we will introduce and discuss different ML-based flow classification techniques according to the types of machine learning paradigms, and followed by our insights at the end.

\subsubsection{Supervised learning-based flow classification}
Supervised learning can achieve higher accuracy of classification among applications. Despite of the tedious labeling work, many supervised learning algorithms have been applied in flow classification, including decision trees, random forests, KNN, and SVM. Trois et al. \cite{trois2018exploring} generated different image textures for different applications, and they classified the flow matrix information using supervised learning algorithms, such as SVM and random forests. Zhao et al. \cite{zhao2019traffic} applied supervised learning algorithms to propose a new classification model that achieved an accuracy of about 99\% in a large supercomputing center.

\subsubsection{Unsupervised learning-based flow classification}
Unsupervised learning-based flow classification techniques do not require labeled datasets, eliminating the difficulties encountered in supervised learning and providing higher robustness. In contrast to supervised learning, the clusters constructed by unsupervised learning need to be mapped to the corresponding applications. However, the large gap between the number of clusters and applications makes it more challenging to classify flows. As investigated in the work of Yan et al. \cite{yan2018survey}, many existing flow classification schemes have adopted unsupervised learning algorithms \cite{baer2016dbstream,bernaille2006traffic,erman2006traffic,wang2017noise,zhang2014robust}. Xiao et al. \cite{xiao2016traffic} focused on the imbalance characteristics of elephant and mice flows in DCNs and proposed a flow classification method using spectral analysis and clustering algorithms. Saber et al. \cite{saber2020online} had a similar research concern and proposed a cost-sensitive classification method that can effectively reduce classification latency. Deque-torres et al. \cite{duque2019approach} proposed a knowledge-defined networking (KDN) based approach for identifying heavy hitters in data center networks, where the efficient threshold for the heavy hitter detection was determined through clustering analysis. Unfortunately, the scheme was not compared with other intelligent methods, thus failing in proving its superiority.

\subsubsection{Deep learning-based flow classification}
The service data and traffic data generated in the data center networks are typically massive, multidimensional, and interrelated. It's very challenging to explore the valuable relationship between these data. To this end, deep learning (e.g., CNN, RNN and LSTM) is introduced to DCN as a promising way to find the potential relationship between these massive and interrelated data. Compared with the former two ML-based classification techniques, the deep learning based schemes have no advantage in training time and classification speed. To this end, Wang et al. \cite{wang2019neural} focused on the speed of classification and implemented a high-speed online flow classifier via field programmable gate array (FPGA), where the authors claimed that it can guarantee an accuracy of more than 99\% while reducing the training time to be one-thousandth of the CPU-based approach. Liu et al. \cite{liu2020fine} implemented a more fine-grained flow classification method based on GRU and reduced flow monitoring costs. In addition, Zeng et al. \cite{zeng2019deep} proposed a lightweight end-to-end framework for flow classification and intrusion detection by deeply integrating flow classification and network security.

\subsubsection{Reinforcement learning-based flow classification}
Reinforcement learning agent iteratively interacts with the environment aiming to find a global optimal classification scheme according to the feedback reward and punishment of feedback in a network scenario. To handle the highly dynamic network conditions in DCNs, Tang et al. \cite{tang2019flow} proposed a new reinforcement learning-based flow splitter that effectively reduced the average completion time of flows, especially for delay-sensitive mice flows. Whereas, as the reinforcement learning tends to fall into local optimal solution and takes a longer training time, this paradigm has not been widely used in traffic classification.

% Flow_Classification_Architecture
\begin{figure}
    \centering
    \includegraphics[width=0.985\linewidth]{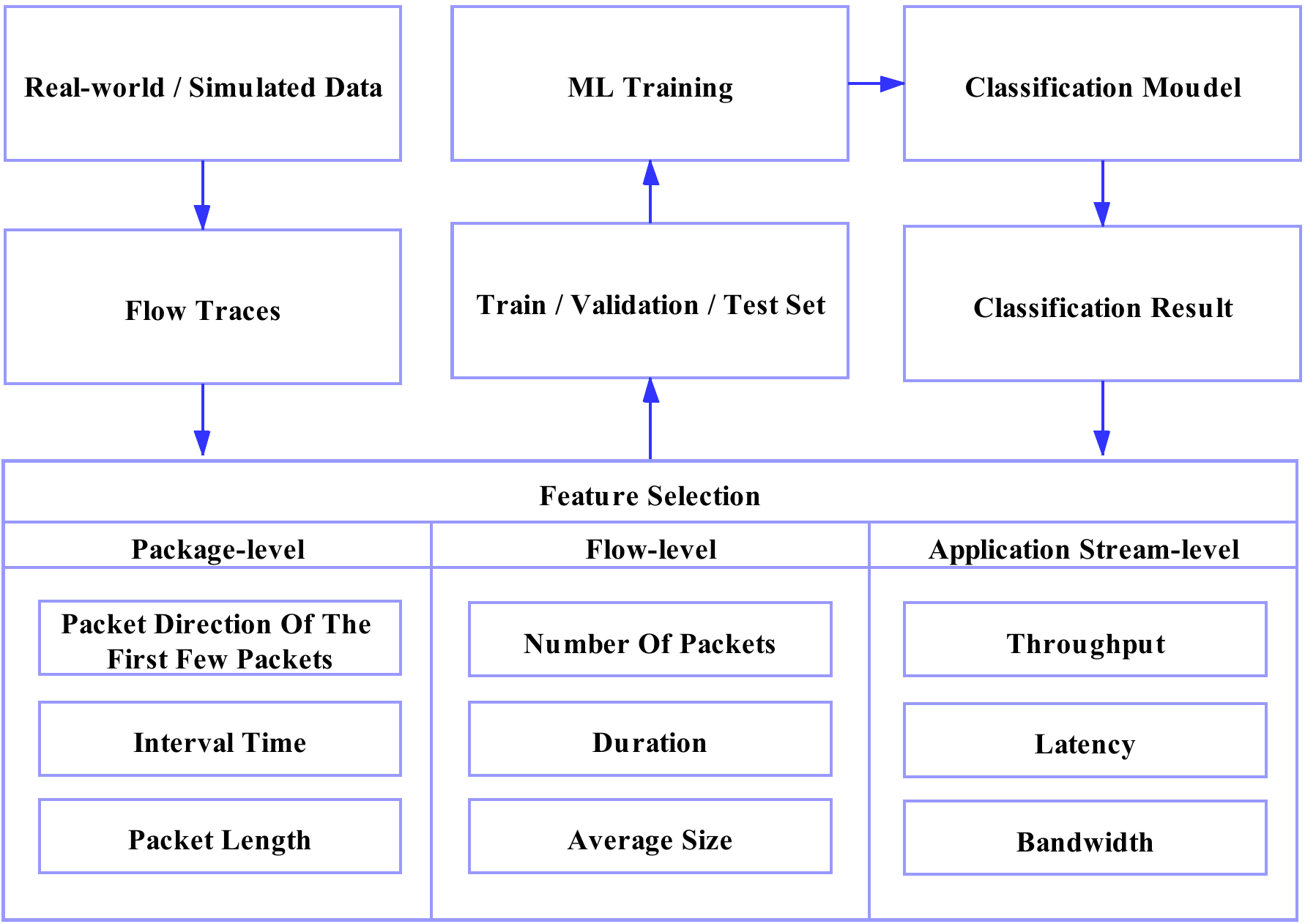}
    \caption{The general system workflow of flow classification, ranging from different levels of feature collections, data pre-processing, model training, to the final model inference outputting classification results}
    \label{Flow_Classification_Architecture}
    \vspace{-0.3cm}
\end{figure}

\subsubsection{Discussion and Insights}
The ML can overcome the limitations and constraints of traditional flow classification schemes. In view of this, numerous ML-based intelligent traffic classification schemes have been proposed. Table \ref{flow_classification} summarizes and compares these existing work from various perspectives. Through systematic investigations and in-depth analysis, in this paper we summarize a general flow classification workflow, as shown in Figure \ref{Flow_Classification_Architecture}, ranging from different levels of feature collections, data pre-processing, model training, to model inference outputting classification results. Furthermore, we have summarized several key concerns that need to be addressed, as listed below.

\begin{itemize}
    \item \textbf{Fine granularity.} The complex diverse DCN service scenarios, high requirements on flow control, and more precise network management are driving the flow classification techniques toward a more fine-grained direction. A fine-grained and accurate classification scheme can allocate network resources efficiently, ensuring a better user experience. However, most of the traditional and intelligent classification schemes are only based on a rough general classification scale, for example, singly based on the network protocol or a single function of the application, which can not provide better QoS. In some simplified scenarios, even if the fine-grained classification has been achieved, the computational overhead, monitoring overhead, stability and feasibility also need to be considered.
    \item \textbf{Flexibility and robustness.} To meet various service needs, flow classification should consider the timeliness and effectiveness of classification, which could help services meet their SLAs. Using FPGA is a feasible way to improve the response speed of classification and avoid the influence of abnormal conditions on the classification efficiency. When encountering the common network anomalies such as jitter, packet loss, and retransmission, the efficiency of a robust flow classification solution should not degrade. Moreover, the quality of extracted data features can also significantly affect the final result of classification, and redundant features will reduce the accuracy of the algorithm along with additional computational overhead \cite{shi2017efficient}.
\end{itemize}

%  Flow Classification
\begin{sidewaystable*}[thp]
    \caption{Research Progress of Data Center Network Intelligence: Flow Classification} % title name of the table
    \label{flow_classification}
    \centering % centering table
    \begin{threeparttable}
        \resizebox{ \linewidth}{!}{
            \begin{tabular}{p{4cm} p{2.7cm} p{8cm} p{6cm}  p{3.5cm} p{3.5cm} p{2cm}<{\centering}} % creating 8 columns
                \specialrule{0em}{1pt}{1pt}
                \hline\hline % inserting double-line
                \textbf{Ref}                             & \textbf{ML Category \& Model Adopted}                                                                                                                                                   & \textbf{Features} & \textbf{Data Source} & \textbf{Additional Constraints} & \textbf{Estimation Function} & \textbf{Experimental comparison subjects} \\
                \specialrule{0em}{1pt}{1pt}
                \hline % inserts single-line
                \specialrule{0em}{1pt}{1pt}
                % Entering 1st row 
                Zeng et al. \cite{zeng2019deep}
                                                         & DL, CNN/LSTM
                                                         & A lightweight end-to-end framework for flow classification and intrusion detection was proposed, and it had excellent performance on two public data sets
                                                         & ISCX 2012 IDS dataset \cite{shiravi2012toward} and ISCX VPN-nonVPN flow dataset \cite{draper2016characterization}
                                                         & Storage
                                                         & $F_1$-score, etc.
                                                         & \makecell[c]{\XSolidBrush  |  \XSolidBrush  |  \CheckmarkBold}

                \\
                \specialrule{0em}{1pt}{1pt}
                \hline % inserts single-line
                \specialrule{0em}{1pt}{1pt}
                % Entering 2st row
                Xiao et al. \cite{xiao2016traffic}       & UL, Clustering Algorithm
                                                         & A fast and effective flow classification method was proposed.
                                                         & Real-world historical data \cite{mawi}
                                                         & Statistical cost
                                                         & Class accuracy  and class recall
                                                         & \XSolidBrush  |  \XSolidBrush |  \CheckmarkBold
                \\
                \specialrule{0em}{1pt}{1pt}
                \hline % inserts single-line
                \specialrule{0em}{1pt}{1pt}
                % Entering 3st row
                Liu et al. \cite{liu2017adaptive}
                                                         & FTR, Unspecified
                                                         & Adopted dynamical flow learning (DTL) algorithm, weighted optimization based on Gaussian distribution and smooth mechanism based on difference estimation.
                                                         & CAIDA \cite{analysiscaida}
                                                         & None
                                                         & Customized
                                                         & \CheckmarkBold  |  \XSolidBrush  |  \CheckmarkBold                                                                                                                                                                                                                                                                                              \\
                \specialrule{0em}{1pt}{1pt}
                \hline % inserts single-line
                \specialrule{0em}{1pt}{1pt}
                % Entering 4st row
                Duque-Torres et al. \cite{duque2019approach}
                                                         & UL, Clustering Algorithm
                                                         & A new intelligent system based on KDN
                                                         & Real-world historical data from University DCN
                                                         & None
                                                         & Silhouette score
                                                         & \CheckmarkBold  |  \XSolidBrush  |  \XSolidBrush
                \\
                \specialrule{0em}{1pt}{1pt}
                \hline % inserts single-line
                \specialrule{0em}{1pt}{1pt}
                % Entering 5st row
                Cuzzocrea et al. \cite{cuzzocrea2015coarse}
                                                         & FTR, KNN, etc.
                                                         & ML classifiers based on different workload models produced the best classifiers
                                                         & Simulated data
                                                         & Workload
                                                         & Accuracy
                                                         & \CheckmarkBold  |  \XSolidBrush  |  \CheckmarkBold
                \\
                \specialrule{0em}{1pt}{1pt}
                \hline % inserts single-line
                \specialrule{0em}{1pt}{1pt}
                % Entering 6st row
                Shekhawat et al. \cite{shekhawat2018datacenter}
                                                         & FTR, SVM etc
                                                         & A method for classifying and characterizing data center workloads based on resource usage was proposed
                                                         & Google Cluster Trace (GCT) \cite{reiss2012towards} and Bit Brains Trace (BBT) \cite{shen2015statistical}
                                                         & Workload
                                                         & SSE
                                                         & \CheckmarkBold  |  \XSolidBrush  |  \CheckmarkBold
                \\
                \specialrule{0em}{1pt}{1pt}
                \hline % inserts single-line
                \specialrule{0em}{1pt}{1pt}
                % Entering 7st row
                Zhu et al. \cite{zhu2020differentiated}
                                                         & DL, GRU
                                                         & Differentiated transmission control services
                                                         & UNB CIC VPN-nonVPN  dataset \cite{draper2016characterization}
                                                         & Buffer size, FCT and throughput
                                                         & $F_1$-score etc
                                                         & \CheckmarkBold  |  \CheckmarkBold  |  \CheckmarkBold
                \\
                \specialrule{0em}{1pt}{1pt}
                \hline % inserts single-line
                \specialrule{0em}{1pt}{1pt}
                % Entering 8st row
                Trois et al. \cite{trois2018exploring}   & SL, SVM/RF
                                                         & Flow forecasting based on flow matrices
                                                         & gathered from MapReduce \cite{dean2008mapreduce} and DCNs
                                                         & None
                                                         & Confusion matrix etc
                                                         & \XSolidBrush   |  \XSolidBrush   |  \CheckmarkBold
                \\
                \specialrule{0em}{1pt}{1pt}
                \hline % inserts single-line
                \specialrule{0em}{1pt}{1pt}
                % Entering 9st row
                Liu et al. \cite{liu2020fine}            & DL, GRU/RF
                                                         & Fine-grained approach to flow classification
                                                         & Real-world historical flow data from a data center \cite{benson2010network}
                                                         & Monitoring cost
                                                         & $F_1$-score etc
                                                         & \CheckmarkBold     | \XSolidBrush   |  \CheckmarkBold
                \\
                \specialrule{0em}{1pt}{1pt}
                \hline % inserts single-line
                \specialrule{0em}{1pt}{1pt}
                % Entering 10st row
                Tang et al. \cite{tang2019flow}          & RL, DDPG
                                                         & Proposed high-performance stream scheduling policy DDPG-FS based on DDPG
                                                         & Simulated data
                                                         & Load, FCT
                                                         & FCT etc
                                                         & \CheckmarkBold     | \CheckmarkBold   |  \CheckmarkBold
                \\
                \specialrule{0em}{1pt}{1pt}
                \hline % inserts single-line
                \specialrule{0em}{1pt}{1pt}
                % Entering 11st row
                Viljoen et al. \cite{viljoen2016machine} & DL, DNN
                                                         & A low overhead, adaptive traffic classifier was proposed
                                                         & Historical flow data from a data center \cite{benson2010network}
                                                         & None
                                                         & Confusion matrix etc
                                                         & \XSolidBrush     |  \XSolidBrush   |  \CheckmarkBold                                                                                                                                                                                                                                                                                            \\
                \specialrule{0em}{1pt}{1pt}
                \hline % inserts single-line
                \specialrule{0em}{1pt}{1pt}
                % Entering 12st row
                Zhu et al. \cite{zhu2019machine}         & DL/UL, LSTM/BIRCH
                                                         & The unsupervised hierarchical clustering algorithm was used to classify the unexecuted jobs, and Davies-Bouldin and other indicators were used to evaluate the clustering quality
                                                         & ClusterData2011\_2 \cite{reiss2011google}, which is a data set released by Google Data Center
                                                         & Workload
                                                         & MSE, etc.
                                                         & \CheckmarkBold     |  \XSolidBrush   |  \CheckmarkBold                                                                                                                                                                                                                                                                                          \\
                \specialrule{0em}{1pt}{1pt}
                \hline % inserts single-line
                \specialrule{0em}{1pt}{1pt}
                % Entering 13st row
                Amaral et al. \cite{amaral2016machine}   & SL, RF
                                                         & The traffic data was collected by Openflow protocol and analyzed by machine learning algorithms such as RF
                                                         & Real-world historical data from unknown sources
                                                         & None
                                                         & Accuracy
                                                         & \CheckmarkBold     |  \XSolidBrush   |  \CheckmarkBold
                \\
                \specialrule{0em}{1pt}{1pt}
                \hline % inserts single-line
                \specialrule{0em}{1pt}{1pt}
                % Entering 14st row
                Bolodurina et al. \cite{bolodurina2017model}
                                                         & Unspecified, Unspecified
                                                         & Accelerated learning of the initial phase of analysis performance, improve flow classification accuracy
                                                         & Simulated data
                                                         & Network Latency
                                                         & Delay of application request
                                                         & \CheckmarkBold     |  \XSolidBrush   |  \CheckmarkBold
                \\
                \specialrule{0em}{1pt}{1pt}
                \hline % inserts single-line
                \specialrule{0em}{1pt}{1pt}
                % Entering 15st row
                Wang et al. \cite{wang2019neural}
                                                         & DL, DNN
                                                         & Implementing a flow classifier using FPGAs with far better performance than CPU
                                                         & Real-world historical data from unknown sources
                                                         & None
                                                         & Training time and accuracy
                                                         & \CheckmarkBold     |  \CheckmarkBold    |  \XSolidBrush
                \\
                \specialrule{0em}{1pt}{1pt}
                \hline % inserts single-line
                \specialrule{0em}{1pt}{1pt}
                % Entering 16st row
                Sarber et al. \cite{saber2020online}
                                                         & SL/UL, BRF/RKNN
                                                         & A cost-sensitive classification method was proposed to guarantee performance while alleviating the data imbalance problem
                                                         & CAIDA \cite{analysispassive}, two university datacenters datasets\cite{theophilus} and an internet flow dataset (UNIBs \cite{unibs})
                                                         & Network resources and latency
                                                         & $F_1$-score, etc.
                                                         & \CheckmarkBold     |  \CheckmarkBold    |  \CheckmarkBold
                \\
                \specialrule{0em}{1pt}{1pt}
                \hline % inserts single-line
                \specialrule{0em}{1pt}{1pt}
                % Entering 17st row
                Wang et al. \cite{wang2018scheduling}
                                                         & SL, C4.5/NBD
                                                         & A packet optical switching network architecture and a priority-aware scheduling algorithm were proposed
                                                         & Simulated data
                                                         & Bandwidth
                                                         & Recall and classification speed
                                                         & \CheckmarkBold     |  \CheckmarkBold    |  \XSolidBrush
                \\
                \specialrule{0em}{1pt}{1pt}
                \hline % inserts single-line
                \specialrule{0em}{1pt}{1pt}
                % Entering 18st row
                Zhao et al. \cite{zhao2019traffic}
                                                         & SL, RF/C4.5/KNN
                                                         & Proposed a clustering flow label propagation technique based on equivalent flow-labeled propagation and a synthetic-flow feature generation algorithm based on Bidirectional-flow (BDF)
                                                         & Real-world historical data from unknown sources
                                                         & None
                                                         & Precision, recall, $F_1$-score  and accuracy
                                                         & \CheckmarkBold     |  \XSolidBrush    |  \CheckmarkBold
                \\
                \hline\hline % inserting double-line
            \end{tabular}
        }
    \end{threeparttable}
\end{sidewaystable*}

%  Assessment of Load Balancing Schemes based on REBEL-3S
\begin{table*}[thp]
    \caption{Assessment of Load Balancing Schemes based on REBEL-3S} % title name of the table
    \label{Assessment of Load Balancing Schemes based on REBEL-3S}
    \centering % centering table
    \begin{threeparttable}
        \resizebox{ \linewidth}{!}{
            \begin{tabular}{lcccccccccc} % creating 9 columns
                \specialrule{0em}{1pt}{1pt}
                \hline\hline % inserting double-line
                \textbf{Ref} & \textbf{Reliability} & \textbf{Energy Efficiency} & \textbf{Bandwidth Utilization} & \textbf{Latency} & \textbf{Security} & \textbf{Stability} & \textbf{Scalability} \\
                \hline % inserts single-line
                \specialrule{0em}{1pt}{1pt}
                % Entering 1st row 
                Ruelas et al. \cite{ruelas2018load}
                             & NO
                             & NO
                             & YES
                             & YES
                             & NO
                             & NO
                             & YES
                \\
                % Entering 2st row 
                Tosounidis et al. \cite{tosounidis2020deep}
                             & NO
                             & NO
                             & YES
                             & YES
                             & NO
                             & YES
                             & YES
                \\
                % Entering 3st row 
                Doke et al. \cite{doke2018deep}
                             & NO
                             & NO
                             & YES
                             & YES
                             & NO
                             & NO
                             & NO
                \\
                % Entering 4st row 
                Hashemi et al. \cite{hashemi2018end}
                             & YES
                             & NO
                             & YES
                             & YES
                             & NO
                             & YES
                             & NO
                \\
                % Entering 5st row 
                Zhou et al. \cite{zhou2019fast}
                             & NO
                             & NO
                             & YES
                             & YES
                             & NO
                             & YES
                             & NO
                \\
                % Entering 6st row 
                Tang et al. \cite{tang2019flow}
                             & NO
                             & NO
                             & YES
                             & YES
                             & NO
                             & YES
                             & YES
                \\
                % Entering 7st row 
                Liu et al. \cite{liu2019intelligent,liu2021drl}
                             & NO
                             & NO
                             & YES
                             & YES
                             & NO
                             & YES
                             & YES
                \\
                % Entering 8st row 
                Yu et al. \cite{yu2019long}
                             & YES
                             & NO
                             & YES
                             & YES
                             & NO
                             & YES
                             & YES
                \\
                % Entering 9st row 
                Zhao et al. \cite{zhao2020machine}
                             & NO
                             & NO
                             & YES
                             & YES
                             & NO
                             & YES
                             & YES
                \\
                % Entering 10st row 
                Liu et al. \cite{liu2019mix}
                             & NO
                             & NO
                             & YES
                             & NO
                             & NO
                             & YES
                             & YES
                \\
                % Entering 11st row 
                Francois et al. \cite{francois2016optimizing}
                             & NO
                             & NO
                             & YES
                             & YES
                             & YES
                             & YES
                             & YES
                \\
                % Entering 12st row 
                Scherer et al. \cite{scherer2015practise}
                             & NO
                             & NO
                             & YES
                             & YES
                             & NO
                             & NO
                             & NO
                \\
                % Entering 13st row 
                Prevost et al. \cite{prevost2011prediction}
                             & NO
                             & YES
                             & NO
                             & NO
                             & NO
                             & NO
                             & NO
                \\
                % Entering 14st row 
                Sun et al. \cite{sun2020qos,sun2020smartfct}
                             & NO
                             & YES
                             & YES
                             & YES
                             & NO
                             & YES
                             & YES
                \\
                % Entering 15st row 
                Lin et al. \cite{lin2018rilnet}
                             & NO
                             & NO
                             & YES
                             & YES
                             & NO
                             & YES
                             & NO
                \\
                % Entering 16st row 
                Wang et al. \cite{wang2018scheduling}
                             & NO
                             & YES
                             & YES
                             & YES
                             & NO
                             & YES
                             & YES
                \\
                % Entering 17st row 
                Li et al. \cite{li2020traffic}
                             & NO
                             & NO
                             & YES
                             & YES
                             & NO
                             & NO
                             & YES
                \\
                \hline\hline % inserting double-line
                \specialrule{0em}{1pt}{1pt}
                \specialrule{0em}{1pt}{1pt}
            \end{tabular}
        }
    \end{threeparttable}
\end{table*}

\subsection{Load Balancing}
The purpose of load balancing is to ensure the balanced distribution of flows on different network routing paths, so as to minimize flow completion time, improve bandwidth utilization and reduce latency. The load balancing problem is usually formulated as a multi-commodity flow (MCF) problem, which has been proved to be NP-hard. The traffic in data centers usually changes in milliseconds or even microseconds, however, traditional unintelligent solutions lack the flexibility of dynamic adjustment according to the real-time network environment status, which may lead to imbalanced load distribution or even network congestion \cite{zhang2011modeling}. As for the performance evaluation and effectiveness verification, there are a variety of metrics. Generally speaking, solutions are usually evaluated in terms of the average link utilization, scalability, robustness, and energy efficiency, which is consistent with the evaluation dimension of REBEL-3S.

\subsubsection{Traditional solutions}
Empirically, the decision-making of load balancing largely depends on  the real-time collected network running status data. According to the way of data acquisition, traditional unintelligent solutions can be divided into two categories: centralized and distributed. Centralized solutions, such as DENS \cite{kliazovich2013dens}, Hedera \cite{al2010hedera}, and Mahout \cite{curtis2011mahout} make decisions based on the global network knowledge acquired through a centralized controller. However, the centralized schemes typically inevitably result in additional communication overhead between the controller and the data plane, which poses extra traffic burden on the network. Besides, the centralized schemes usually require dedicated and customized hardware (e.g. OpenFlow supported), which are cross vendor incompatible. Distributed solution is difficult to make the best decision without a global view. Although the topology of DCNs is often symmetrical in design, it is still difficult to deal with the network failure caused by the damage to hardware devices. Alizadeh et al. \cite{alizadeh2014conga} insisted that an efficient load balancing scheme must address the asymmetry issue caused by network failures which are highly disruptive. Importantly, traditional approaches are difficult to learn from the historical traffic data and automatically adjust the strategies to achieve network optimization.

\subsubsection{Machine learning-based solutions}
Facing the ever-changing network environment, ML can help the network self-learning, realize the self-decision of flow scheduling strategy, and self-adaptation to the network environment. Zhao et al. \cite{zhao2020machine} proposed two ML-assisted flow aggregation schemes to achieve low latency and high bandwidth. They improved network throughput through specifically designed optical cross-connect switches, and deployed ML algorithms (such as DT, KNN, and SVM) with relaxed accuracy requirements to edge nodes to reduce latency. The Wavelength Division Multiplexing (WDM) technology was used to improve the scalability of the optical network, but the FPGA board was installed on each ToR to perform feature sampling, increasing the hardware cost. Wang et al. \cite{wang2018scheduling} used supervised learning algorithms, such as C4.5, to classify network flows with different characteristics and developed a priority-aware scheduling algorithm for packet switching. The simulation experiments showed that their scheduling algorithm was superior to the classical RR algorithm \cite{hahne1991round} with respect to the average delay and packet loss rates. Compared with the former two kinds of learning paradigms, deep learning algorithms have better applicability and more and more researchers prefer to use them. Li et al. \cite{li2020traffic} designed a GNN-based optimizer for flow scheduling to reduce the flow completion time (FCT). However, GNN brings a more complex network structure and increases computational costs. Prevost et al. \cite{prevost2011prediction} devoted to the energy consumption problem caused by the imbalanced load. They proposed a new DNN-based framework to achieve load demand prediction and stochastic state transfer. With the increasing difficulty of optimization goal, more and more researchers consider using reinforcement learning and deep reinforcement learning to deal with the dynamic network environment. Tang et al. \cite{tang2019flow} employed a modified DDPG algorithm for high-performance flow scheduling. Compared to the native DDPG and traditional unintelligent methods, their solution significantly reduced the FCT of delay-sensitive flows.

\subsubsection{Discussion and Insights}
To avoid the difficulty of collecting real-world data, 90\% of the intelligent solutions were tested based on simulation-generated data. Due to the diversity and complexity of DCN application scenarios and the differences in data sources and scenarios, it is not easy to make a fair comparison between intelligent schemes. As a result, over 76\% of the solutions lack comparisons to other ML-based solutions. Besides, inheriting the advantages of traditional unintelligent solutions, more than 40\% of intelligent solutions adopt SDN architecture to collect network data and make decisions based on a centralized controller.

The details of the existing intelligent solutions are listed in Table \ref{Flow_Scheduling_and_Load_Balancing}, and the assessment results of each solution based on REBEL-3S are summarized in Table \ref{Assessment of Load Balancing Schemes based on REBEL-3S}. Clearly, most solutions consider bandwidth utilization and latency, account for 94\% and 88\% respectively, while few solutions take security and reliability into account, account for 6\% and 12\% respectively. In addition to the issues and challenges discussed above, the following two concerns need to be considered.

\begin{itemize}
    \item \textbf{Compatibility of network stacks.} According to the work of Wang et al. \cite{wang2020improving}, one of the necessary conditions for many traditional research work is to be compatible with different network protocol stacks \cite{alizadeh2012less,alizadeh2013pfabric,wilson2011better,hong2012finishing,bai2017pias,chen2016scheduling}. With the advent of new network protocols such as $D^2$TCP \cite{vamanan2012deadline} and DCTCP \cite{alizadeh2010data}, the unequal distribution of network bandwidth among different network users due to stack incompatibility has attracted considerable attention \cite{he2016ac,cronkite2016virtualized}. Intelligent solutions should focus on compatibility between different network protocols and prevent unfair resource allocation caused by the different protocol parameters.
    \item \textbf{Dynamic of network flow.} In respect of the dynamic of network traffic, it is necessary and beneficial to adjust the threshold and priority of flows in time. It is suggested to pay attention to the local and overall benefits, especially for the mixed flow scheduling problem in the scheduling process.
\end{itemize}

% Load Balancing
\begin{sidewaystable*}[thp]
    \caption{Research Progress of Data Center Network Intelligence: Load Balancing} % title name of the table
    \label{Flow_Scheduling_and_Load_Balancing}
    \centering % centering table
    \begin{threeparttable}
        \resizebox{ \linewidth}{!}{
            \begin{tabular}{p{3.5cm} p{2.5cm} p{8.5cm} p{4cm} p{4cm} p{4cm} p{2.3cm} p{3.5cm}<{\centering}} % creating 8 columns
                \hline\hline % inserting double-line
                \textbf{Ref} & \textbf{ML Category \& Model Adopted}                                                                                                                                                                                                       & \textbf{Features} & \textbf{Data Source} & \textbf{Feature Selection} & \textbf{Additional Constraints} & \textbf{Estimation Function} & \textbf{Experimental comparison subjects} \\
                \specialrule{0em}{1pt}{1pt}
                \hline % inserts single-line
                \specialrule{0em}{1pt}{1pt}
                % Entering 1st row 
                Ruelas et al. \cite{ruelas2018load}
                             & DL, DNN
                             & Based on the ideas of KDN and DNN, a new load balancing method was provided
                             & Simulated data
                             & Bandwidth, Latency
                             & None
                             & MSE
                             & \CheckmarkBold  |  \XSolidBrush  |  \XSolidBrush
                \\
                \specialrule{0em}{1pt}{1pt}
                \hline % inserts single-line
                \specialrule{0em}{1pt}{1pt}
                % Entering 2st row 
                Tosounidis et al. \cite{tosounidis2020deep}
                             & DRL, DQN
                             & Deep reinforcement learning was used to efficiently load balance service requests in DCNs
                             & Simulated data
                             & Bandwidth/CPU/Memory utilization
                             & CPU computing power
                             & RTT, etc.
                             & \CheckmarkBold  |  \CheckmarkBold  |  \XSolidBrush
                \\
                \specialrule{0em}{1pt}{1pt}
                \hline % inserts single-line
                \specialrule{0em}{1pt}{1pt}
                % Entering 3st row 
                Doke et al. \cite{doke2018deep}
                             & DRL, DQN
                             & Try to apply DQN algorithm to network traffic management
                             & Simulated data
                             & Number of session requests
                             & None
                             & Standard deviation
                             & \CheckmarkBold  |  \CheckmarkBold  |  \XSolidBrush
                \\
                \specialrule{0em}{1pt}{1pt}
                \hline % inserts single-line
                \specialrule{0em}{1pt}{1pt}
                % Entering 4st row 
                Hashemi et al. \cite{hashemi2018end}
                             & DL, CNN
                             & An end-to-end real-time flow management system based on deep learning was proposed
                             & Simulated data
                             & Time-varying flow sizes
                             & Accidents and adverse weather conditions, etc.
                             & AvgRMSE, etc.
                             & \CheckmarkBold  |  \CheckmarkBold  |  \XSolidBrush
                \\
                \specialrule{0em}{1pt}{1pt}
                \hline % inserts single-line
                \specialrule{0em}{1pt}{1pt}
                % Entering 5st row 
                Zhou et al. \cite{zhou2019fast}
                             & DL, RNN
                             & A residual flow compression mechanism was introduced to minimize the completion time of data-intensive applications
                             & Simulated data
                             & Coflow width, coflow size and arrival time, etc.
                             & Network bandwidth
                             & FCT, etc.
                             & \CheckmarkBold  |  \CheckmarkBold  |  \XSolidBrush
                \\
                \specialrule{0em}{1pt}{1pt}
                \hline % inserts single-line
                \specialrule{0em}{1pt}{1pt}
                % Entering 6st row 
                Tang et al. \cite{tang2019flow}
                             & RL, DDPG
                             & Proposed high-performance stream scheduling policy DDPG-FS based on DDPG
                             & Simulated data
                             & Maximum operational flow between links and maximum total flow demand, etc.
                             & Load
                             & AvgFCT, etc.
                             & \CheckmarkBold     |  \CheckmarkBold   |  \CheckmarkBold
                \\
                \specialrule{0em}{1pt}{1pt}
                \hline % inserts single-line
                \specialrule{0em}{1pt}{1pt}
                % Entering 7st row 
                Liu et al. \cite{liu2021drl,liu2019intelligent}
                             & DRL, DQN/DDPG
                             & The DRL-R (Deep Reinforcement Learning-based Routing) algorithm was proposed to bridge multiple resources (node cache, link bandwidth) by quantifying the contribution of multiple resources (node cache, link bandwidth) to reduce latency
                             & Simulated data
                             & Node cache, Link bandwidth
                             & Maximize overall network throughput while meeting QoS
                             & FCT, etc.
                             & \CheckmarkBold  |  \CheckmarkBold  |  \CheckmarkBold
                \\
                \specialrule{0em}{1pt}{1pt}
                \hline % inserts single-line
                \specialrule{0em}{1pt}{1pt}
                % Entering 8st row 
                Yu et al. \cite{yu2019long}
                             & DL, BRNN/LSTM
                             & A flow scheduling method capable of extracting long-term flow characteristics
                             & Historical flow data obtained from three university data centers
                             & 5-tuple, etc.
                             & Network Resource Utilization
                             & MAE, MRE, and RMSE, etc.
                             & \CheckmarkBold  |  \XSolidBrush  |  \CheckmarkBold
                \\
                \specialrule{0em}{1pt}{1pt}
                \hline % inserts single-line
                \specialrule{0em}{1pt}{1pt}
                % Entering 9st row 
                Zhao et al. \cite{zhao2020machine}
                             & SL/UL, SVM/DT, etc.
                             & Proposed two ML-assisted traffuc aggregation schemes that can effectively improve throughput, reduce latency and FCT
                             & Simulated data
                             & Network throughput, network latency and flow completion time
                             & Transaction response processing time
                             & Accuracy, etc.
                             & \CheckmarkBold  |  \CheckmarkBold  |  \XSolidBrush
                \\
                \specialrule{0em}{1pt}{1pt}
                \hline % inserts single-line
                \specialrule{0em}{1pt}{1pt}
                % Entering 10st row 
                Liu et al. \cite{liu2019mix}
                             & DRL, DDPG/CNN
                             & A hybrid flow scheduling scheme based on deep reinforcement learning was proposed
                             & Simulated data
                             & Paths and flows information
                             & Maximizing deadline satisfaction rate for mice flows and minimizing FCT for elephant flows
                             & Deadline meet rate \cite{liu2017information}
                             & \CheckmarkBold  |  \CheckmarkBold  |  \XSolidBrush
                \\
                \specialrule{0em}{1pt}{1pt}
                \hline % inserts single-line
                \specialrule{0em}{1pt}{1pt}
                % Entering 11st row 
                Francois et al. \cite{francois2016optimizing}
                             & DL/RL, RNN
                             & A logical centralized cognitive routing engine was developed based on stochastic NNs with reinforcement learning
                             & 5 geographically dispersed data centers
                             & Latency
                             & None
                             & RTT, etc.
                             & \CheckmarkBold  |  \CheckmarkBold  |  \XSolidBrush
                \\
                \specialrule{0em}{1pt}{1pt}
                \hline % inserts single-line
                \specialrule{0em}{1pt}{1pt}
                % Entering 12st row 
                Scherer et al. \cite{scherer2015practise}
                             & DL, DNN
                             & A neural network-based framework for server workload prediction was proposed
                             & VM workload tracking and real-time data from real-time systems recorded in private cloud data centers operated by IBM
                             & CPU, memory, disk, and network
                             & None
                             & Accuracy
                             & \CheckmarkBold  |  \XSolidBrush  |  \CheckmarkBold
                \\
                \specialrule{0em}{1pt}{1pt}
                \hline % inserts single-line
                \specialrule{0em}{1pt}{1pt}
                % Entering 13st row 
                Prevost et al. \cite{prevost2011prediction}
                             & DL, DNN
                             & A new framework combining load demand forecasting and stochastic state transfer models was proposed to minimize energy consumption while maintaining network performance
                             & URL resource requests for NASA's web server and EPA's web server
                             & Unspecified
                             & Minimize energy consumption while guaranteeing performance
                             & RMSE, RMSE
                             & \CheckmarkBold  |  \CheckmarkBold  |  \XSolidBrush
                \\
                \specialrule{0em}{1pt}{1pt}
                \hline % inserts single-line
                \specialrule{0em}{1pt}{1pt}
                % Entering 14st row 
                Sun et al. \cite{sun2020qos,sun2020smartfct}
                             & DRL, DDPG
                             & The DRL algorithm was used to improve power efficiency and ensure FCT
                             & Wikipedia trace files \cite{urdaneta2009wikipedia}
                             & Number of switch ports, port power, incoming and outgoing flow rates, etc.
                             & QoS and energy consumption
                             & FCT, etc.
                             & \CheckmarkBold  |  \CheckmarkBold  |  \XSolidBrush
                \\
                \specialrule{0em}{1pt}{1pt}
                \hline % inserts single-line
                \specialrule{0em}{1pt}{1pt}
                % Entering 15st row 
                Lin et al. \cite{lin2018rilnet}
                             & DRL, DDPG
                             & Reinforcement-based learning was used to learn a network and perform load balancing by aggregating flows
                             & Simulated data
                             & Throughput between nodes
                             & Network Overheads
                             & Customized
                             & \CheckmarkBold  |  \CheckmarkBold  |  \XSolidBrush
                \\
                \specialrule{0em}{1pt}{1pt}
                \hline % inserts single-line
                \specialrule{0em}{1pt}{1pt}
                % Entering 16st row 
                Wang et al. \cite{wang2018scheduling}
                             & SL, C4.5/NBD
                             & A packet-switched optical network (PSON) architecture and a priority-aware scheduling algorithm were proposed
                             & Simulated data
                             & Virtual queue priority, number of packets, latency, etc.
                             & Bandwidth
                             & Recall and classification speed
                             & \CheckmarkBold  |  \CheckmarkBold  |  \XSolidBrush
                \\
                \specialrule{0em}{1pt}{1pt}
                \hline % inserts single-line
                \specialrule{0em}{1pt}{1pt}
                % Entering 17st row 
                Li et al. \cite{li2020traffic}
                             & DL, GNN
                             & To be able to support relational reasoning and combinatorial generalization, the authors have proposed a GNN-based flow optimization method
                             & Simulated data
                             & 5-tuple, bandwidth
                             & None
                             & FCT, etc.
                             & \CheckmarkBold  |  \CheckmarkBold  |  \XSolidBrush
                \\
                \hline\hline % inserting double-line
            \end{tabular}
        }
    \end{threeparttable}
\end{sidewaystable*}

\subsection{Resource Management}
As one of the most critical optimization problems in data center, resource management involves the allocation, scheduling, and optimization of computing, storage, network and other resources, which directly affects the overall resource utilization efficiency and resource availability of data center, and further affects the user experience and the revenue of service providers. However, with the increasing complexity of network infrastructure, the explosive growth of the number of hardware devices, and the growing demand for services, the traditional unintelligent solutions can no longer effectively deal with these problems, and there is an urgent need for some intelligent resource management solutions. Studies reveal that ML-assisted intelligent resource allocation can maximize the profit of service providers, provide better quality of experience (QoE) for tenants, and effectively reduce energy costs.

There has been a wide variety of resource management solutions, such as multi-level queues \cite{karthick2014efficient}, simulated annealing \cite{pandit2014resource}, priority-based \cite{soni2014novel}, and heuristic algorithms \cite{bey2015new}. The advent of virtualization allows virtual machines (VMs), virtual containers (VCs), and virtual networks (VNs) to be implemented on a shared physical server. Whereas, the association between various hardware resources (such as CPU, memory and disk space) and virtual resources is highly dynamic throughout the life cycle of services, which is difficult to grasp clearly. The preliminary research findings demonstrated that traditional unintelligent resource management methods can not mine the potential relationships between complex parameters quickly and dynamically. Besides, multi-objective optimization also increases the difficulty of network optimization, such as considering QoS, energy cost and performance optimization at the same time. Furthermore, in a large-scale data center, the complex configuration is also a challenging and destructive problem, where once the configuration error occurs, it will cause incalculable damage to network services, especially for latency-sensitive services. ML can make up for the deficiency of traditional unintelligent methods by learning historical data to dynamically make appropriate management strategies adaptively. Therefore, many researchers have begun to study in this direction, and the solution combined with machine learning came into being. The work of Fiala and Joe \cite{fiala2015survey} explored the application of ML techniques for resource management in the cloud computing area, and Murali et al. \cite{murali2018machine} focused on a distributed NN-based ML approach to achieve efficient resource allocation.

In data center networks, the types of network resources are rich and diverse. At the network level, it can be a physical hardware resource (server, switch, port, link, CPU, memory) or an abstract software resource (virtual network, virtual node, virtual link, virtual switch). In addition, network resources can be task/job-oriented or QoS-oriented. From the perspective of the resource life cycle, resource management can also focus on resource prediction or resource utilization optimization. In view of the diversity of resource management methods and the difference of optimization objectives, we divide ML-based resource management schemes into the following five categories: task-oriented, virtual entity-oriented, QoS-oriented, resource prediction-oriented, and resource utilization-oriented resource management.

\subsubsection{Task-oriented Resource Management}
In data centers, there are various special tasks with different particular performance requirements, such as compute-intensive tasks and latency-sensitive tasks, which require that the resource management solutions can be customized for different tasks. Tesauro et al. \cite{tesauro2006hybrid} used reinforcement learning to optimize the allocation of computing resources by global arbitration, allocated efficient server resources (such as bandwidth and memory) for each web application, and solved the limitations of reinforcement learning by queuing model policy. Marahatta et al. \cite{marahatta2020pefs} classified tasks into failure-prone and failure-prone tasks by DNN and executed different allocation policies for different types of tasks. In addition, resource management can also help reduce the energy consumption of the data center. Yi et al. \cite{yi2020efficient} scheduled the load of computationally intensive tasks based on deep reinforcement learning to minimize the energy cost.

\subsubsection{Virtual Entities-oriented Resource Management}
Virtualization allows tasks with different service and performance requirements to share a series of resources. Generally speaking, virtualized entities include virtual machines (VMs), virtual containers (VCs), and virtual networks (VNs), and we define the solutions that allocate resources for virtualized entities as virtual entities-oriented resource allocation management.

In order to ensure network performance while minimizing power consumption, Caviglione et al. \cite{caviglione2020deep} applied a DRL algorithm, named Rainbow DQN, to solve the multi-objective VM placement problem. Their model was based on the percentages of network capacity, CPU, and disk, with full consideration of energy cost, network security, and QoS. Liu et al. \cite{liu2017hierarchical} applied Q-learning algorithm to distributed management of resources, and their proposed hierarchical network architecture can provide resource allocation and power management of VMs. Experiments showed that when the physical server clusters are set to 30, for 95000 jobs, the proposed hierarchical framework can reduce the network energy consumption and latency by 16.12\% and 16.67\% respectively, compared with the DRL-based resource allocation. It can be seen that in addition to the resource allocation for tasks, the resource allocation for virtual entities also greatly affects the operation efficiency and power consumption of the data center. Elprince et al. \cite{elprince2013autonomous} designed a dynamic resource allocator, which allocated resources through different machine learning techniques (such as REPTree and Linear Regression), and dynamically adjusted the allocated resources through a resource fuzzy tuner. Experiments showed that their solutions can guarantee SLA well and meet differentiated service requirements between various customers. Jobava et al. \cite{jobava2018achieving} managed VM resources through flow-aware consolidation. The AL algorithm was used to divide the virtual clusters to reduce the total communication cost, and then simulated annealing algorithm was employed for intelligent allocation of VM clusters. Both phases were traffic aware.

\subsubsection{QoS-oriented Resource Management}
Resource management optimization research has improved the overall QoS of the service by optimizing resource allocation, although this is not the primary key objective. The two typical representative research work aiming at QoS are as follows. Wang et al. \cite{wang2020case} proposed an on-demand resource scheduling method based on DNN to ensure the QoS of delay-sensitive applications. Wadwadkar et al. \cite{yadwadkar2018machine} leveraged SVM to perform resource prediction and meet QoS requirements with a performance-aware resource allocation policy. As for the uncertainty of prediction, they introduced the concept of confidence measure to mitigate this problem.

% Resource Management
\begin{sidewaystable*}[thp]
    \caption{Research Progress of Data Center Network Intelligence: Resource Management} % title name of the table
    \label{Resource_Management_1}
    \centering % centering table
    \begin{threeparttable}
        \resizebox{ \linewidth}{!}{
            \begin{tabular}{p{3.5cm}  p{1.5cm}<{\centering} p{2.5cm} p{4cm} p{4cm} p{4.5cm} p{4.5cm} p{4.5cm} p{2cm}<{\centering}} % creating 8 columns
                \hline\hline % inserting double-line
                \textbf{Ref} & \textbf{Category \tnote{1}}                                                                                                                                       & \textbf{ML Category \& Model Adopted} & \textbf{Features} & \textbf{Data Source} & \textbf{Feature Selection} & \textbf{Additional Constraints} & \textbf{Estimation Function} & \textbf{Experimental comparison subjects} \\
                \specialrule{0em}{1pt}{1pt}
                \hline % inserts single-line
                \specialrule{0em}{1pt}{1pt}
                % Entering 1st row 
                Wang et al. \cite{wang2020case}
                             & Q
                             & DL, DNN
                             & Monitor the performance of applications in real time, and adjust the resource allocation policy of the corresponding app if performance is impaired to ensure QoS
                             & A web search engine from TailBench benchmark
                             & Internal and external performance counter data
                             & QoS
                             & CPU Quota, etc.
                             & \CheckmarkBold  |  \CheckmarkBold  |  \XSolidBrush
                \\
                \specialrule{0em}{1pt}{1pt}
                \hline % inserts single-line
                \specialrule{0em}{1pt}{1pt}
                % Entering 2st row 
                Che et al. \cite{che2020deep}
                             & RU
                             & DRL, Actor-critic
                             & Task Scheduling, with the goal of optimizing resource utilization and task completion time
                             & Alibaba Cluster Trace Program \cite{2021alibaba}
                             & Task and VM Properties
                             & None
                             & Average task latency time, average task priority, task congestion level
                             & \CheckmarkBold  |  \CheckmarkBold  |  \CheckmarkBold
                \\
                \specialrule{0em}{1pt}{1pt}
                \hline % inserts single-line
                \specialrule{0em}{1pt}{1pt}
                % Entering 3st row 
                Liu et al. \cite{liu2017hierarchical}
                             & V
                             & DRL, Q-learning
                             & Using a hierarchical network architecture for VM resource allocation and power management
                             & Real-world data center workload traces from Google cluster-usage traces \cite{2021google}
                             & Task arrival time, task duration, task resource requests (CPU, memory, disk requirements)
                             & Minimize energy consumption and maintain reasonable performance
                             & Average delay, etc.
                             & \CheckmarkBold  |  \CheckmarkBold  |  \CheckmarkBold
                \\
                \specialrule{0em}{1pt}{1pt}
                \hline % inserts single-line
                \specialrule{0em}{1pt}{1pt}
                % Entering 4st row 
                Tesauro et al. \cite{tesauro2006hybrid}
                             & T
                             & DRL, DNN/Sarsa
                             & Allocate server resources to applications
                             & Simulated data
                             & Application and server properties
                             & None
                             & SLA Total Revenue, etc.
                             & \CheckmarkBold  |  \CheckmarkBold  |  \XSolidBrush
                \\
                \specialrule{0em}{1pt}{1pt}
                \hline % inserts single-line
                \specialrule{0em}{1pt}{1pt}
                % Entering 5st row 
                Liu et al. \cite{liu2019learning,liu2018learning}
                             & T
                             & DRL, DQN
                             & Reduce service latency with data placement policies
                             & MSR Cambridge Traces \cite{narayanan2008write}
                             & End-to-end node information, read/write latency, subsequent analysis latency, etc.
                             & Network Latency
                             & Average data read latency, etc.
                             & \CheckmarkBold  |  \CheckmarkBold  |  \CheckmarkBold
                \\
                \specialrule{0em}{1pt}{1pt}
                \hline % inserts single-line
                \specialrule{0em}{1pt}{1pt}
                % Entering 6st row 
                Yang et al. \cite{yang2020reinforcement}
                             & RU
                             & RL, Q-learning
                             & The storage mode of resources was studied to realize the joint optimization of data storage and traffic management
                             & Simulated data
                             & Transfer rate, remaining server capacity, etc.
                             & None
                             & Transmission completion time, total number of dropped packets, etc.
                             & \CheckmarkBold  |  \CheckmarkBold  |  \XSolidBrush
                \\
                \specialrule{0em}{1pt}{1pt}
                \hline % inserts single-line
                \specialrule{0em}{1pt}{1pt}
                % Entering 7st row 
                Iqbal et al. \cite{iqbal2020adaptive}
                             & RP
                             & DL, DNN
                             & Predicting resource utilization and proposing a resource estimation model
                             & Three publicly available datasets \cite{2021alibaba,gwat13,gwat12}
                             & VM CPU, memory, network and disk utilization.
                             & None
                             & MSE
                             & \CheckmarkBold  |  \CheckmarkBold  |  \CheckmarkBold
                \\
                \specialrule{0em}{1pt}{1pt}
                \hline % inserts single-line
                \specialrule{0em}{1pt}{1pt}
                % Entering 8st row 
                Elprince et al. \cite{elprince2013autonomous}
                             & V
                             & SL/UL, REPTree/Linear Regression, etc.
                             & Allocate resources to VCs based on ML
                             & Real-world traces of Los Alamos National Lab
                             & Detailed information about resource requests and usage, such as memory and CPU time
                             & Guaranteed SLA
                             & RMSE, RAE, RRSE, etc.
                             & \CheckmarkBold  |  \XSolidBrush  |  \CheckmarkBold
                \\
                \specialrule{0em}{1pt}{1pt}
                \hline % inserts single-line
                \specialrule{0em}{1pt}{1pt}
                % Entering 9st row 
                Caviglione et al. \cite{caviglione2020deep}
                             & V
                             & DRL, DQN
                             & Used DRL to solve multi-target VM placement problem
                             & Self-built data center
                             & Percentage of CPU, disk and network requested by VM, etc.
                             & Energy overhead, QoS, network security
                             & Customized
                             & \CheckmarkBold  |  \CheckmarkBold  |  \XSolidBrush
                \\
                \specialrule{0em}{1pt}{1pt}
                \hline % inserts single-line
                \specialrule{0em}{1pt}{1pt}
                % Entering 10st row 
                Yi et al. \cite{yi2020efficient}
                             & T
                             & DRL, DQN
                             & Allocate server resources to applications
                             & Simulated data
                             & Processor utilization, temperature, power consumption, number of spare cores
                             & Minimize power consumption and maintain reasonable performance
                             & Processor temperature and power consumption, etc.
                             & \CheckmarkBold  |  \CheckmarkBold  |  \CheckmarkBold
                \\
                \hline\hline % inserting double-line
            \end{tabular}
        }
        \begin{tablenotes}
            \footnotesize
            \item [1] For convenience, we use "T" for Task-oriented Resource Management, "V" for Virtual Entities-oriented Resource Management, "Q" for QoS-oriented Resource Management, "RP" for Resource\\ Prediction-oriented Resource Management, and "RU" for Resource Utilization-oriented Resource Management.
        \end{tablenotes}
    \end{threeparttable}
\end{sidewaystable*}

% Resource Management
\begin{sidewaystable*}[thp]
    \caption{Research Progress of Data Center Network Intelligence: Resource Management} % title name of the table
    \label{Resource_Management_2}
    \centering % centering table
    \begin{threeparttable}
        \resizebox{ \linewidth}{!}{
            \begin{tabular}{p{3.5cm}  p{1.5cm}<{\centering} p{2.5cm} p{4cm} p{4cm} p{4.5cm} p{4.5cm} p{4.5cm} p{2cm}<{\centering}} %  creating 8 columns
                \hline\hline % inserting double-line
                \textbf{Ref} & \textbf{Category \tnote{1}}                                                                                                                                                   & \textbf{ML Category \& Model Adopted} & \textbf{Features} & \textbf{Data Source} & \textbf{Feature Selection} & \textbf{Additional Constraints} & \textbf{Estimation Function} & \textbf{Experimental comparison subjects} \\
                \specialrule{0em}{1pt}{1pt}
                \hline % inserts single-line
                \specialrule{0em}{1pt}{1pt}
                % Entering 11st row 
                Li et al. \cite{ll2013empowering}
                             & RU
                             & SL, M5P\cite{hall2009weka}
                             & Modeling training on resources / QoS / Workload based on historical data to help develop better scheduling algorithms to better balance throughput, QoS and energy efficiency
                             & Realistic workloads and environments
                             & Average computation time per request and average number of bytes exchanged per request, etc.
                             & Energy overhead, QoS
                             & Service level agreement fulfillment, etc.
                             & \CheckmarkBold  |  \CheckmarkBold  |  \CheckmarkBold
                \\
                \specialrule{0em}{1pt}{1pt}
                \hline % inserts single-line
                \specialrule{0em}{1pt}{1pt}
                % Entering 12st row 
                Thonglek et al. \cite{thonglek2019improving}
                             & RP
                             & DL, LSTM
                             & Predicting job requirements
                             & Google dataset \cite{more}
                             & Requested and used CPU, memory resources
                             & None
                             & CPU and memory utilization
                             & \CheckmarkBold  |  \XSolidBrush  |  \CheckmarkBold
                \\
                \specialrule{0em}{1pt}{1pt}
                \hline % inserts single-line
                \specialrule{0em}{1pt}{1pt}
                % Entering 13st row 
                Xu et al. \cite{xu2017intelligent}
                             & RU
                             & RL, Unspecified
                             & Blockchain-based resource provisioning
                             & Google dataset \cite{liu2016quantitative}
                             & Request resources needed for migration, such as CPU cores, RAM, Disk, etc.
                             & Energy overhead
                             & Energy cost, etc.
                             & \CheckmarkBold  |  \CheckmarkBold  |  \XSolidBrush
                \\
                \specialrule{0em}{1pt}{1pt}
                \hline % inserts single-line
                \specialrule{0em}{1pt}{1pt}
                % Entering 14st row 
                Zerwas et al. \cite{zerwas2019ismael}
                             & RP
                             & SL/UL, CNN/LR, etc.
                             & An ML framework for predicting virtual cluster acceptance rates
                             & Simulated data
                             & Link capacity and free resources, etc.
                             & None
                             & Calculation of unit utilization, etc.
                             & \CheckmarkBold  |  \CheckmarkBold  |  \CheckmarkBold
                \\
                \specialrule{0em}{1pt}{1pt}
                \hline % inserts single-line
                \specialrule{0em}{1pt}{1pt}
                % Entering 15st row 
                Chen et al. \cite{chen2019learning}
                             & T
                             & RL, Actor-critic
                             & Allocate resources to jobs (model-free RL), minimizing task latency
                             & Google dataset
                             & Assignment completion time, waiting time, etc.
                             & Network Latency
                             & Normalized latency, etc.
                             & \CheckmarkBold  |  \CheckmarkBold  |  \CheckmarkBold
                \\
                \specialrule{0em}{1pt}{1pt}
                \hline % inserts single-line
                \specialrule{0em}{1pt}{1pt}
                % Entering 16st row 
                Yu et al. \cite{yu2018leveraging}
                             & RP
                             & DL, DNN
                             & A deep learning-based resource allocation algorithm
                             & Self-built data center
                             & Resource usage status and prioritization of existing flow
                             & None
                             & Path blocking probability, etc.
                             & \CheckmarkBold  |  \CheckmarkBold  |  \CheckmarkBold
                \\
                \specialrule{0em}{1pt}{1pt}
                \hline % inserts single-line
                \specialrule{0em}{1pt}{1pt}
                % Entering 17st row 
                Yadwadkar et al. \cite{yadwadkar2018machine}
                             & Q
                             & SL, SVM
                             & Multi-dimensional solutions
                             & Multiple Hadoop deployments (including deployments for Facebook and Cloudera customers)
                             & CPU utilization, disk utilization and other information
                             & QoS
                             & Accuracy, etc.
                             & \CheckmarkBold  |  \CheckmarkBold  |  \XSolidBrush
                \\
                \specialrule{0em}{1pt}{1pt}
                \hline % inserts single-line
                \specialrule{0em}{1pt}{1pt}
                % Entering 18st row 
                Telenyk et al. \cite{telenyk2018modeling}
                             & RU
                             & RL, Q-learning
                             & Operating with Global Manager for VM scheduling, VM consolidation, etc. for resource optimization, energy saving
                             & Bitbrains \cite{weerasiri2017taxonomy}
                             & CPU, memory and network bandwidth
                             & Energy overhead, QoS
                             & Number of SLA violations, etc.
                             & \CheckmarkBold  |  \CheckmarkBold  |  \XSolidBrush
                \\
                \specialrule{0em}{1pt}{1pt}
                \hline % inserts single-line
                \specialrule{0em}{1pt}{1pt}
                % Entering 19st row 
                Jobava et al. \cite{jobava2018achieving}
                             & V
                             & RL, AL
                             & VM resource management through flow-aware consolidation
                             & Simulated data
                             & Cost matrix
                             & None
                             & Total cost of communication, etc.
                             & \CheckmarkBold  |  \XSolidBrush  |  \XSolidBrush
                \\

                \specialrule{0em}{1pt}{1pt}
                \hline % inserts single-line
                \specialrule{0em}{1pt}{1pt}
                % Entering 20st row 
                Marahatta et al. \cite{marahatta2020pefs}
                             & T
                             & RL, DNN
                             & Allocate resources for tasks (predict tasks based on ML, and divide the tasks into error-prone and non-error-prone tasks, and implement different allocation strategies)
                             & Eular dataset and Internet dataset
                             & The requested resources of the tasks, the actual allocation of resources and whether failure occurred
                             & Energy overhead, QoS
                             & Task failure rate, etc.
                             & \CheckmarkBold  |  \CheckmarkBold  |  \XSolidBrush
                \\
                \specialrule{0em}{1pt}{1pt}
                \hline % inserts single-line
                \specialrule{0em}{1pt}{1pt}
                % Entering 21st row 
                Wang et al. \cite{wang2016presto,wang2015efficient}
                             & V
                             & FTR, BI
                             & The first to apply BI paradigm to solve the VNE problem
                             & Simulated data
                             & CPU capacity and bandwidth capacity between nodes, etc.
                             & Minimize embedding costs and improve economic revenue
                             & Customized
                             & \CheckmarkBold  |  \CheckmarkBold  |  \XSolidBrush
                \\
                \specialrule{0em}{1pt}{1pt}
                \hline % inserts single-line
                \specialrule{0em}{1pt}{1pt}
                % Entering 22st row 
                Rayan et al. \cite{rayan2018resource}
                             & RP
                             & SL, SVM, etc.
                             & Workload prediction based on ML, based on which the required resources can be predicted
                             & Simulated data
                             & Number of active physical machines and power consumption data
                             & Energy overhead
                             & RMSE, running time
                             & \CheckmarkBold  |  \CheckmarkBold  |  \CheckmarkBold
                \\
                \hline\hline % inserting double-line
            \end{tabular}
        }
        \begin{tablenotes}
            \footnotesize
            \item [1] For convenience, we use "T" for Task-oriented Resource Management, "V" for Virtual Entities-oriented Resource Management, "Q" for QoS-oriented Resource Management, "RP" for Resource\\ Prediction-oriented Resource Management, and "RU" for Resource Utilization-oriented Resource Management.
        \end{tablenotes}
    \end{threeparttable}
\end{sidewaystable*}

% Assessment of Resource Management Schemes based on REBEL-3S
\begin{table*}[thp]
    \caption{Assessment of Resource Management Schemes based on REBEL-3S} % title name of the table
    \label{Assessment of Resource Management Schemes based on REBEL-3S}
    \centering % centering table
    \resizebox{ \linewidth}{!}{
        \begin{tabular}{lcccccccccc}
            \hline\hline % inserting double-line
            \textbf{Ref} & \textbf{Reliability} & \textbf{Energy Efficiency} & \textbf{Bandwidth Utilization} & \textbf{Latency} & \textbf{Security} & \textbf{Stability} & \textbf{Scalability} \\
            \hline % inserts single-line
            \specialrule{0em}{1pt}{1pt}
            % Entering 1st row 
            Wang et al. \cite{wang2020case}
                         & YES
                         & NO
                         & YES
                         & YES
                         & NO
                         & YES
                         & NO
            \\
            % Entering 2st row 
            Che et al. \cite{che2020deep}
                         & NO
                         & NO
                         & NO
                         & YES
                         & NO
                         & YES
                         & NO
            \\
            % Entering 3st row 
            Liu et al. \cite{liu2017hierarchical}
                         & NO
                         & YES
                         & YES
                         & YES
                         & NO
                         & YES
                         & NO
            \\
            % Entering 4st row 
            Tesauro et al. \cite{tesauro2006hybrid}
                         & NO
                         & NO
                         & YES
                         & YES
                         & NO
                         & YES
                         & NO
            \\
            % Entering 5st row 
            Liu et al. \cite{liu2018learning,liu2019learning}
                         & YES
                         & YES
                         & YES
                         & YES
                         & NO
                         & YES
                         & YES
            \\
            % Entering 6st row 
            Yang et al. \cite{yang2020reinforcement}
                         & NO
                         & NO
                         & YES
                         & YES
                         & NO
                         & YES
                         & YES
            \\
            % Entering 7st row 
            Iqbal et al. \cite{iqbal2020adaptive}
                         & NO
                         & NO
                         & YES
                         & YES
                         & NO
                         & NO
                         & NO
            \\
            % Entering 8st row 
            Elprince et al. \cite{elprince2013autonomous}
                         & NO
                         & NO
                         & YES
                         & YES
                         & NO
                         & NO
                         & YES
            \\
            % Entering 9st row 
            Caviglione et al. \cite{caviglione2020deep}
                         & NO
                         & YES
                         & YES
                         & YES
                         & YES
                         & YES
                         & NO
            \\
            % Entering 10st row 
            Yi et al. \cite{yi2020efficient}
                         & NO
                         & YES
                         & YES
                         & YES
                         & NO
                         & YES
                         & YES
            \\
            % Entering 11st row 
            Li et al. \cite{ll2013empowering}
                         & NO
                         & YES
                         & YES
                         & NO
                         & NO
                         & NO
                         & NO
            \\
            % Entering 12st row 
            Thonglek et al. \cite{thonglek2019improving}
                         & NO
                         & NO
                         & YES
                         & NO
                         & NO
                         & YES
                         & NO
            \\
            % Entering 13st row 
            Xu et al. \cite{xu2017intelligent}
                         & NO
                         & YES
                         & YES
                         & NO
                         & NO
                         & YES
                         & NO
            \\
            % Entering 14st row 
            Zerwas et al. \cite{zerwas2019ismael}
                         & NO
                         & YES
                         & YES
                         & YES
                         & NO
                         & NO
                         & NO
            \\
            % Entering 15st row 
            Chen et al. \cite{chen2019learning}
                         & NO
                         & NO
                         & YES
                         & YES
                         & NO
                         & YES
                         & NO
            \\
            % Entering 16st row 
            Yu et al. \cite{yu2018leveraging}
                         & NO
                         & NO
                         & YES
                         & YES
                         & NO
                         & NO
                         & NO
            \\
            % Entering 17st row 
            Yadwadkar et al. \cite{yadwadkar2018machine}
                         & YES
                         & YES
                         & YES
                         & YES
                         & NO
                         & YES
                         & YES
            \\
            % Entering 18st row 
            Telenyk et al. \cite{telenyk2018modeling}
                         & NO
                         & YES
                         & YES
                         & YES
                         & NO
                         & YES
                         & NO
            \\
            % Entering 19st row 
            Jobava et al. \cite{jobava2018achieving}
                         & YES
                         & YES
                         & YES
                         & YES
                         & NO
                         & YES
                         & YES
            \\
            % Entering 20st row 
            Marahatta et al. \cite{marahatta2020pefs}
                         & YES
                         & YES
                         & YES
                         & NO
                         & NO
                         & YES
                         & NO
            \\
            % Entering 21st row 
            Wang et al. \cite{wang2016presto,wang2015efficient}
                         & YES
                         & YES
                         & YES
                         & NO
                         & NO
                         & NO
                         & YES
            \\
            % Entering 22st row 
            Rayan et al. \cite{rayan2018resource}
                         & NO
                         & YES
                         & YES
                         & NO
                         & NO
                         & NO
                         & NO
            \\
            \hline\hline % inserting double-line
            \specialrule{0em}{1pt}{1pt}
            \specialrule{0em}{1pt}{1pt}
        \end{tabular}
    }
\end{table*}

\subsubsection{Resource Prediction-oriented Resource Management}
Resource prediction plays an essential role in resource management. Timely and accurate resource forecasting can make the data center achieve more effective resource scheduling, and further improve the overall performance of the data center network. However, although virtualization and other technologies greatly enrich the types of resources and improve service efficiency, it also increases the difficulty of resource prediction. Besides, Aguado et al. \cite{aguado2016towards} implied that the prediction accuracy of traditional unintelligent algorithms cannot be guaranteed on account of diversity of services and bandwidth explosion. Moreover, to cope with the unpredictable resource demand, traditional resource management mechanisms usually over-allocate resources to ensure the availability of resources, which is harmful to the overall resource utilization of data centers. How to deal with the differentiated requirements of various workloads and precisely predict resources still remains a challenge. Yu et al. \cite{yu2018leveraging} proposed a deep learning-based flow prediction and resource allocation strategy in optical DCNs, and experimental results demonstrated that their approach achieved a better performance compared with a single-layer NN-based algorithm. Iqbal et al. \cite{iqbal2020adaptive} proposed an adaptive observation window resizing method based on a 4-hidden-layer DNN for resource utilization estimation. The work of Thonglek et al. \cite{thonglek2019improving} predicted the required resources for jobs by a two-layer LSTM network, which outperformed the traditional RNN model, with improvements of 10.71\% and 47.36\% in CPU and memory utilization, respectively.

\subsubsection{Resource Utilization-oriented Resource Management}
Resource utilization is regarded as an intuitive and important metric to evaluate a resource management mechanism. This type of resource management schemes typically improve the resource utilization through task scheduling, VM migration and load balancing algorithms. It is worth noting that dynamic change of resource demand in data centers requires the algorithm being able to automatically optimize resource utilization according to the changing network environment. However, the traditional unintelligent solutions are difficult to cope with the high variability of the network environment. Therefore, a few researchers have begun to apply machine learning to solve these problems. Che et al. performed task scheduling based on the actor-critic deep reinforcement learning algorithm to optimize resource utilization and task completion time \cite{che2020deep}. Telenyk et al. \cite{telenyk2018modeling} used the Q-learning algorithm for global resource management, and realized resource optimization and energy saving through virtual machine scheduling and virtual machine aggregation. In addition to improving resource utilization through scheduling and consolidation, Yang et al. \cite{yang2020reinforcement} focused their research on the optimization of storage resource, that is, how to efficiently store data. They used distributed multi-agent reinforcement learning methods to achieve joint optimization of resources, which effectively improved network throughput and reduced stream transmission time.

\subsubsection{Discussion and Insights}
The current resource management system in today's data centers is complex and multifaceted. Along with the expansion of service scenarios, the resource scheduling among various virtualized entities is getting more complicated. Increasingly, researchers adopt deep learning or deep reinforcement learning aiming to achieve a more intelligent resource management. We list the details of each intelligent solution in Tables \ref{Resource_Management_1} and \ref{Resource_Management_2}. Then, we evaluate these solutions with respect to each dimension of REBEL-3S, as shown in Table \ref{Assessment of Resource Management Schemes based on REBEL-3S}. It reveals that more than half of the solutions take the energy efficiency into account in resource management \cite{caviglione2020deep,ll2013empowering,telenyk2018modeling}, and most of them consider the impact of network stability. Here, we summarize several key concerns, as below, which deserve to be further studied and addressed.

\begin{itemize}
    \item \textbf{Stability and scalability of models.} Taking reinforcement learning as an example, primary decisions may have relatively poor consequences due to a lack of domain knowledge or good heuristic strategies \cite{tesauro2006hybrid}. When the agent performs tentative actions, it may fall into local optimal solutions if not appropriately trained. Besides, reinforcement learning may lack good scalability in large DCNs.
    \item \textbf{Adaptability to Multi-objective and multi-task.} Whether it is a traditional resource allocation scheme (such as priority-based VM allocation \cite{soni2014novel}, heuristic-based resource allocation \cite{bey2015new}), or an intelligent resource allocation scheme, their performance is usually evaluated in a specific single scenario. Whereas, one qualified intelligent solution should fully consider the richness of scenarios and requirements, and be able to adapt to multi-scenario and multi-task network environment.
    \item \textbf{Security of resource allocation.} The flexibility of virtualized resources can make vulnerability or fault propagation faster, and fault recovery and fault source tracing more difficult. Padhy et al. \cite{padhy2011cloud} disclosed that vulnerabilities were found in VMware's shared folder mechanism, which could allow the users of guest systems to read and write to any part of the host file system, including system folders and other security-sensitive files.
    \item \textbf{Perspective of resource lifecycle.} The allocation, utilization, and recycling of resources occur frequently. Current intelligent solutions focus more on the prediction of resource allocation and maximization of benefits in the process of resource allocation, but lack related studies on resource collection and recycling.
\end{itemize}

\subsection{Routing Optimization}
In DCNs, routing optimization is one of the most important research areas and has aroused some discussions in both academia and industry. With the advantage of SDN, routing optimization can get a global view of the network and deploy strategies conveniently, but the existing traditional SDN-based methods cannot sensitively adapt the real-time traffic changes in data center networks \cite{amezquita2019efficient,wang2018efficient,xiao2017openflow,guo2018balancing,wang2014freeway,liu2014sdn}. For instance, if the routing policies cannot be timely adjusted according to the dynamic network conditions, the imbalance of network flows may cause uneven load distribution among network nodes, where some nodes are highly loaded or even overloaded while some other nodes are underutilized or even idle, resulting in the waste of resources and the degradation of QoS.

The rise of ML techniques has brought new thinking to this field. Chen et al. \cite{chen2018knowledge,chen2018deep} pointed out that ML-based routing schemes can efficiently solve path optimization problems in complex dynamic network environments, while traditional unintelligent routing schemes are difficult in achieving similar results under the same conditions. However, our investigation shows that so far there has been relatively little research in this field, and most of the existing ML-based routing solutions in data centers are centralized schemes based on SDN. In this paper, we divide the small amount of existing ML-based research work into intra-DC and inter-DC routing optimization.

%  A_General_ML-assisted_Routing_Framework
\begin{figure}
    \centering
    \includegraphics[width=\linewidth]{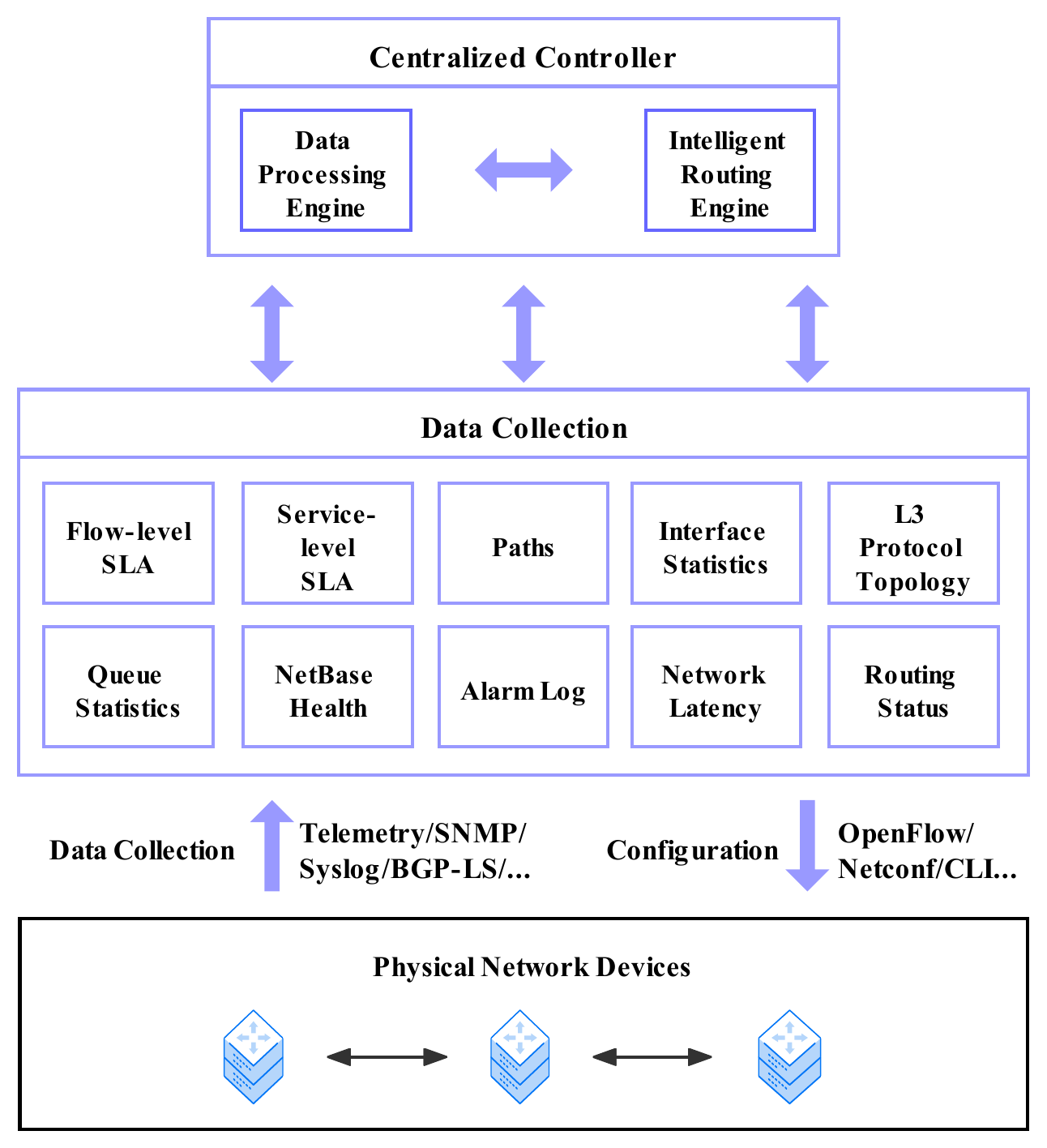}
    \caption{A General ML-assisted Routing Framework}
    \label{A_General_ML-assisted_Routing_Framework}
    \vspace{-0.3cm}
\end{figure}

% Assessment of Routing Optimization Schemes based on REBEL-3S
\begin{table*}[thp]
    \caption{Assessment of Routing Optimization Schemes based on REBEL-3S}
    \label{Assessment of Routing Optimization Schemes based on REBEL-3S}
    \centering % centering table
    \resizebox{ \linewidth}{!}{
        \begin{tabular}{lcccccccccc} % creating 9 columns
            \hline\hline % inserting double-line
            \textbf{Ref} & \textbf{Reliability} & \textbf{Energy Efficiency} & \textbf{Bandwidth Utilization} & \textbf{Latency} & \textbf{Security} & \textbf{Stability} & \textbf{Scalability} \\
            \hline % inserts single-line
            \specialrule{0em}{1pt}{1pt}
            % Entering 1st row 
            Bolodurina et al. \cite{bolodurina2018comprehensive}
                         & NO
                         & NO
                         & YES
                         & YES
                         & NO
                         & YES
                         & NO
            \\
            % Entering 2st row 
            Fu et al. \cite{fu2020deep}
                         & NO
                         & NO
                         & YES
                         & YES
                         & NO
                         & YES
                         & NO
            \\
            % Entering 3st row 
            Yao et al. \cite{yao2020dqn}
                         & NO
                         & YES
                         & YES
                         & YES
                         & NO
                         & YES
                         & NO
            \\
            % Entering 4st row 
            Yu et al. \cite{yu2018drom}
                         & NO
                         & NO
                         & YES
                         & YES
                         & NO
                         & YES
                         & NO
            \\
            % Entering 5st row 
            Panda \cite{panda2020energy}
                         & NO
                         & YES
                         & YES
                         & YES
                         & NO
                         & YES
                         & NO
            \\
            % Entering 6st row 
            Zhou et al. \cite{zhou2019fast}
                         & NO
                         & NO
                         & YES
                         & YES
                         & NO
                         & YES
                         & NO
            \\
            % Entering 7st row 
            Liu et al. \cite{liu2021drl,liu2019intelligent}
                         & NO
                         & NO
                         & YES
                         & YES
                         & NO
                         & YES
                         & YES
            \\
            % Entering 8st row 
            Liu et al. \cite{liu2019mix}
                         & NO
                         & NO
                         & YES
                         & NO
                         & NO
                         & YES
                         & YES
            \\
            % Entering 9st row 
            Hong et al. \cite{hong2020neural}
                         & NO
                         & NO
                         & YES
                         & NO
                         & NO
                         & YES
                         & NO
            \\
            % Entering 10st row 
            Francois et al. \cite{francois2016optimizing}
                         & NO
                         & NO
                         & NO
                         & YES
                         & YES
                         & YES
                         & YES
            \\
            \hline\hline % inserting double-line
            \specialrule{0em}{1pt}{1pt}
            \specialrule{0em}{1pt}{1pt}
        \end{tabular}
    }
\end{table*}

\subsubsection{Intra-DC Routing Optimization}
About 80\% of researchers paid attention to the routing optimization within a data center. Bolodurina et al. proposed \cite{bolodurina2018comprehensive} a routing optimization strategy taking the SLA into consideration. They pre-classified communication channels based on unsupervised learning and deep learning to obtain a detailed network feature set and then performed clustering according to SLA requirements. Finally, the network feature set and clustering data were used as input for NN training to obtain a suitable routing strategy. Likewise, Fu et al. \cite{fu2020deep} also adopted a pre-classification strategy before using deep Q-learning (DQL) to train different NNs for elephant flows and mice flows, respectively. They computed optimal routing paths for different types of flows with the help of SDN, ensuring low latency for mice flows and high throughput for elephant flows. Nevertheless, this solution suffers high computation overhead and requires a relatively long time in path calculation. Yu et al. \cite{yu2018drom} used the DDPG algorithm for routing decisions and improved the network performance providing stable and high-quality routing services. It achieved good convergence and effectiveness while guaranteeing QoS (delay minimization and throughput maximization). Yao et al. \cite{yao2020dqn} designed a DQN-based energy-aware routing algorithm to find energy-efficient data forwarding paths and control paths for switches in data centers.

\subsubsection{Inter-DC Routing Optimization}
Routing optimization among data centers has also received much attention. Hong et al. \cite{hong2020neural} used NN to predict flow load blocking probability and proposed an efficient routing and a resource allocation policy among data centers. Comparatively, Francois et al. \cite{francois2016optimizing} proposed a logically centralized cognitive routing engine (CRE) based on reinforcement learning and deep learning to meet SLAs through a cognitive routing engine. The CRE can work well even in highly chaotic environments, where it can leverage RNN with RL to find the efficient overlay paths with minimal monitoring overhead among geographically dispersed data centers. Panda and Satyasen \cite{panda2020energy} combined the adaptive ant colony optimization algorithm with the neural network to solve the power consumption and routing optimization problems in the elastic optical data center network.

\subsubsection{Discussion and Insights}
We list details of each intelligent solution in Table \ref{Routing_Optimization}, and evaluate each solution from various dimensions according to REBEL-3S in Table \ref{Assessment of Routing Optimization Schemes based on REBEL-3S}. Besides, we summarize a general ML-assisted routing framework based on the investigation results (as shown in Figure \ref{A_General_ML-assisted_Routing_Framework}). The intelligent routing decision module is mainly implemented in the centralized controller. The control plane collects real-time network operation data and performance data through the OpenFlow protocol, and then take advantage of ML to train and calculate the efficient routing policies. Finally, these network policies are distributed to the network devices for execution through the SDN southbound interface. Based on the limited research work in this area, we summarize the following several concerns as well as some potential research opportunities.

\begin{itemize}
    \item \textbf{Multi-objective optimization.} Considering the diversity of traffic types with different priorities, how to provide differentiated routing optimization policies is deemed as a challenging task. What's more, the decision making of routing policies may also depend on the preliminary results of some other optimization models, such as traffic identification and classification schemes, where it will involve collaborative optimizations of multiple learning models for multiple tasks, which further increases the difficulty of optimization. In addition, the optimization objectives are often diverse as well, and even need to be satisfied at the same time (such as high throughput, low latency, high reliability, load balancing, high link utilization, fault tolerance / burst tolerance, and even high energy efficiency, etc.), which poses a great challenge to the optimality of the solutions (global optimal or local optimal) and the computational complexity of algorithms.
    \item \textbf{ML model selection.} Choosing an appropriate and effective ML model is the first and most important step towards intelligent routing optimization. However, in view of the particularity of the data center scenario, in most cases, the existing ML models can not be applied directly to the network routing optimization. For instance, the original Q-learning algorithm, which can only deal with discrete action problems, cannot handle the continuous dynamic changes in the network, and the complex and diverse network states may lead to excessive storage space for Q-table. In consideration of the fact that the research work on ML-based routing optimization is still relatively little, the types of ML models used in the existing schemes are relatively few.
          As of now the investigation shows that deep reinforcement learning is still the mainstream paradigm for solving routing optimization problems. Thereby, it is deemed that more effective ML models need to be explored and validated, which is of great necessity and importance.
\end{itemize}

% Routing Optimization
\begin{sidewaystable*}[thp]
    \caption{Research Progress of Data Center Network Intelligence: Routing Optimization} % title name of the table
    \label{Routing_Optimization}
    \centering % centering table
    \begin{threeparttable}
        \resizebox{ \linewidth}{!}{
            \begin{tabular}{p{3.1cm} p{1.5cm}<{\centering} p{2.5cm} p{5.3cm} p{3cm} p{4cm} p{4cm} p{4cm} p{3.5cm}<{\centering}} % creating 8 columns
                \hline\hline % inserting double-line
                \textbf{Ref} & \textbf{Category \tnote{1}}                                                                                                                                                                                                                 & \textbf{ML Category \& Model Adopted} & \textbf{Features} & \textbf{Data Source} & \textbf{Feature Selection} & \textbf{Additional Constraints} & \textbf{Estimation Function} & \textbf{Experimental comparison subjects} \\
                \specialrule{0em}{1pt}{1pt}
                \hline % inserts single-line
                \specialrule{0em}{1pt}{1pt}
                % Entering 1st row 
                Bolodurina et al. \cite{bolodurina2018comprehensive}
                             & A
                             & UL/DL, DNN
                             & Optimized routing based on network characteristics and QoS requirements
                             & Simulated data
                             & Channel delay, jitter, packet loss rate, QoS rules used, etc.
                             & QoS
                             & Resource capacity and network response time
                             & \CheckmarkBold  |  \CheckmarkBold  |  \XSolidBrush
                \\
                \specialrule{0em}{1pt}{1pt}
                \hline % inserts single-line
                \specialrule{0em}{1pt}{1pt}
                % Entering 2st row 
                Fu et al. \cite{fu2020deep}
                             & A
                             & DRL, DQN/CNN
                             & Efficient routing strategies for elephant and mice flows, respectively
                             & Simulated data
                             & Source switch, destination switch and bandwidth requirements for streams, etc.
                             & Low latency and low packet loss rate for mice streams and high throughput and low packet loss rate for elephant streams
                             & Customized
                             & \CheckmarkBold  |  \CheckmarkBold  |  \XSolidBrush
                \\
                \specialrule{0em}{1pt}{1pt}
                \hline % inserts single-line
                \specialrule{0em}{1pt}{1pt}
                % Entering 3st row 
                Yao et al. \cite{yao2020dqn}
                             & A
                             & DRL, DQN
                             & An energy-efficient load balancing strategy
                             & Simulated data
                             & Source node, destination node and bandwidth requirements of the flow, etc.
                             & Energy consumption
                             & Customized
                             & \CheckmarkBold  |  \CheckmarkBold  |  \XSolidBrush
                \\
                \specialrule{0em}{1pt}{1pt}
                \hline % inserts single-line
                \specialrule{0em}{1pt}{1pt}
                % Entering 4st row 
                Yu et al. \cite{yu2018drom}
                             & A
                             & DRL, DDPG
                             & DDPG-based routing policy with excellent convergence and effectiveness
                             & Simulated data
                             & Latency, throughput, etc.
                             & QoS
                             & Network latency, throughput
                             & \CheckmarkBold  |  \CheckmarkBold  |  \XSolidBrush
                \\
                \specialrule{0em}{1pt}{1pt}
                \hline % inserts single-line
                \specialrule{0em}{1pt}{1pt}
                % Entering 5st row 
                Panda \cite{panda2020energy}
                             & B
                             & DL, DNN
                             & Solving power consumption and routing optimization problems of elastic optical networks based on NNs and adaptive ant colony optimization algorithms
                             & Simulated data
                             & Input weight matrix of encoder, etc.
                             & Energy consumption
                             & Spectrum utilization, etc.
                             & \CheckmarkBold  |  \CheckmarkBold  |  \XSolidBrush
                \\
                \specialrule{0em}{1pt}{1pt}
                \hline % inserts single-line
                \specialrule{0em}{1pt}{1pt}
                % Entering 6st row 
                Zhou et al. \cite{zhou2019fast}
                             & A
                             & DL, RNN
                             & A residual flow compression mechanism was introduced to minimize the completion time of data-intensive applications
                             & Simulated data
                             & Coflow width, coflow size and arrival time, etc.
                             & Network bandwidth
                             & FCT, etc.
                             & \CheckmarkBold  |  \CheckmarkBold  |  \XSolidBrush
                \\
                \specialrule{0em}{1pt}{1pt}
                \hline % inserts single-line
                \specialrule{0em}{1pt}{1pt}
                % Entering 7st row 
                Liu et al. \cite{liu2021drl,liu2019intelligent}
                             & A
                             & DRL, DQN/DDPG
                             & The DRL-R (Deep Reinforcement Learning-based Routing) algorithm was proposed to bridge multiple resources (node cache, link bandwidth) by quantifying the contribution of multiple resources (node cache, link bandwidth) to reduce latency
                             & Simulated data
                             & Node cache, link bandwidth
                             & Maximize overall network throughput while meeting QoS
                             & FCT, etc.
                             & \CheckmarkBold  |  \CheckmarkBold  |  \XSolidBrush
                \\
                \specialrule{0em}{1pt}{1pt}
                \hline % inserts single-line
                \specialrule{0em}{1pt}{1pt}
                % Entering 8st row 
                Liu et al. \cite{liu2019mix}
                             & A
                             & DRL, DDPG/CNN
                             & A hybrid flow scheduling scheme based on deep reinforcement learning was proposed
                             & Simulated data
                             & Paths and flows information
                             & Maximizing deadline satisfaction rate for mice flow and minimizing FCT for elephant flow
                             & Deadline meet rate \cite{liu2017information}
                             & \CheckmarkBold  |  \CheckmarkBold |  \XSolidBrush
                \\
                \specialrule{0em}{1pt}{1pt}
                \hline % inserts single-line
                \specialrule{0em}{1pt}{1pt}
                % Entering 9st row 
                Hong et al. \cite{hong2020neural}
                             & B
                             & DL, DNN
                             & An efficient routing and resource allocation strategy between data centers.
                             & Simulated data
                             & Number of racks, average capacity required for connection requests on racks, etc.
                             & None
                             & CDF
                             & \CheckmarkBold  |  \XSolidBrush  |  \XSolidBrush
                \\
                \specialrule{0em}{1pt}{1pt}
                \hline % inserts single-line
                \specialrule{0em}{1pt}{1pt}
                % Entering 10st row 
                Francois et al. \cite{francois2016optimizing}
                             & B
                             & DL/RL, RNN
                             & A logical centralized cognitive routing engine was developed based on stochastic NNs with reinforcement learning
                             & 5 geographically dispersed data centers
                             & Latency
                             & QoS
                             & RTT etc
                             & \CheckmarkBold  |  \CheckmarkBold  |  \XSolidBrush
                \\
                \hline\hline % inserting double-line
            \end{tabular}
        }
        \begin{tablenotes}
            \footnotesize
            \item[1] For convenience, we use "A" for Intra-DC Routing Optimization and "B" for Inter-DC Routing Optimization.
        \end{tablenotes}
    \end{threeparttable}
\end{sidewaystable*}

% Assessment of Congestion Control Schemes based on REBEL-3S
\begin{table*}[thp]
    \caption{Assessment of Congestion Control Schemes based on REBEL-3S}
    \label{Assessment of Congestion Control Schemes based on REBEL-3S}
    \centering % centering table
    \resizebox{ \linewidth}{!}{
        \begin{tabular}{lcccccccccc} % creating 9 columns
            \hline\hline % inserting double-line
            \textbf{Ref} & \textbf{Reliability} & \textbf{Energy Efficiency} & \textbf{Bandwidth Utilization} & \textbf{Latency} & \textbf{Security} & \textbf{Stability} & \textbf{Scalability} \\
            \hline % inserts single-line
            \specialrule{0em}{1pt}{1pt}
            % Entering 1st row 
            Jin et al. \cite{jin2018congestion}
                         & NO
                         & NO
                         & YES
                         & NO
                         & NO
                         & YES
                         & YES
            \\
            % Entering 2st row 
            Liao et al. \cite{liao2020fast,liao2019deep}
                         & NO
                         & NO
                         & YES
                         & YES
                         & NO
                         & YES
                         & YES
            \\
            % Entering 3st row 
            Liu et al. \cite{liu2019learning,liu2018learning}
                         & YES
                         & YES
                         & YES
                         & YES
                         & NO
                         & YES
                         & YES
            \\
            % Entering 4st row 
            Majidi et al. \cite{majidi2020dc}
                         & NO
                         & YES
                         & YES
                         & YES
                         & NO
                         & YES
                         & YES
            \\
            % Entering 5st row 
            Nie \cite{nie2019dynamic}
                         & NO
                         & NO
                         & YES
                         & YES
                         & NO
                         & YES
                         & YES
            \\
            % Entering 6st row 
            Ruffy et al. \cite{ruffy2018iroko}
                         & NO
                         & NO
                         & YES
                         & YES
                         & NO
                         & YES
                         & YES
            \\
            % Entering 7st row 
            Thiruvenkatam et al. \cite{thiruvenkatam2020optimizing}
                         & NO
                         & NO
                         & YES
                         & YES
                         & NO
                         & YES
                         & YES
            \\
            % Entering 8st row 
            Sun et al. \cite{sun2020qos,sun2020smartfct}
                         & NO
                         & YES
                         & YES
                         & YES
                         & NO
                         & YES
                         & YES
            \\
            % Entering 9st row 
            Rastegarfar et al. \cite{rastegarfar2016tcp}
                         & YES
                         & YES
                         & YES
                         & YES
                         & NO
                         & YES
                         & YES
            \\
            % Entering 10st row 
            Xiao et al. \cite{xiao2019tcp}
                         & YES
                         & NO
                         & YES
                         & YES
                         & NO
                         & YES
                         & YES
            \\
            \hline\hline % inserting double-line
            \specialrule{0em}{1pt}{1pt}
            \specialrule{0em}{1pt}{1pt}
        \end{tabular}
    }
\end{table*}

\subsection{Congestion Control}
\label{Congestion Control}
The complexity and diversity of service scenarios and finer granularity of flow demands have made congestion control more complicated in data centers. For instance, some applications require high micro-burst tolerance \cite{shan2017improving,shan2018micro}, while some applications demand low latency \cite{mittal2015timely} or high throughput \cite{gao2017demepro}. Besides, the diverse applications and computing frameworks with different characteristics in data centers further produce a variety of traffic patterns, such as one-to-one, one-to-many, many-to-one, many-to-many, and all-to-all traffic patterns. However, the traditional TCP-based solutions can hardly meet all the requirements of these different traffic patterns at the same time \cite{flach2013reducing,dong2018pcc}, which often result in queueing delay, jitter incast, throughput collapse, increased flow completion time, and packet loss \cite{choudhury1998dynamic,lu2018dynamic,majidi2019adaptive}.

Admittedly, congestion control (CC) is the core of the TCP algorithm, which determines the data transmission efficiency. Although the research on CC has spanned more than three decades, the vast majority of CC solutions in data center network scenarios have followed a conservative strategy. It starts transmission at a slow sending rate and then uses certain strategies (e.g., AIMD) to adjust the sending rate during subsequent transmissions, which is normally agnostic to the flow deadline and network congestion and cannot well cope with the micro-burst scenario as well \cite{cho2017credit,jose2015high}. For instance, when multiple synchronous servers send data to a single receiver simultaneously, the shallow-buffered switches at the last hop are prone to be overwhelmed by the bursty traffic resulting in increased queueing delay or even packet loss, which is known as the TCP incast problem \cite{sreekumari2016early}. Explicit Congestion Notification (ECN) is the most common congestion handling mechanism, and MQ-ECN \cite{bai2016enabling} is the first protocol to enable multi-queue scenarios in data centers, guaranteeing queue independence to ensure no loss of network latency and throughput. Nevertheless, the traffic is inherently bursty in DCNs, and MQ-ECN is only applicable to round-based scheduling mechanisms.

Recently, ML has attracted researchers' interest, and some ML-based CC algorithms have been proposed. From the perspective of the decision-making mode of CC policies, we divide the existing ML-based solutions into centralized and distributed congestion control.

\subsubsection{Centralized Congestion Control}
The centralized scheme detects, avoids, and mitigates network congestion through unified scheduling and centralized management of decentralized network resources. The centralized allocation of network resources can maximize overall network resource utilization, but it may also pose some problems. The transmission of relevant network logs will take up additional bandwidth and memory, and the relatively long response time of network policy decisions will have an adverse impact on the latency sensitive applications. What's more, in order to achieve efficient congestion control, many solutions require to customize hardware, which has lost the generality and practicality. More importantly, the centralized congestion control scheme typically suffers from the scalability issue, where the centralized controller usually becomes the bottleneck, which is difficult to adapt to large-scale data center network.

Jin et al. \cite{jin2018congestion} designed two congestion control methods based on the improved Q-learning algorithm and Sarsa algorithm, respectively, with the help of a centralized SDN controller. They believed that the congestion control algorithm should consider the temporal and spatial characteristics of the flows and focused on the path of the current flow when selecting the action. However, the improved algorithms were not competent to adapt to complex data center networks \cite{jiang2021machine}. It is necessary to further optimize the design of the reward function (for example, adding delay, power consumption, fairness, reliability, and other factors into the evaluation dimension), and to test under more complex network environment. Ruffy et al. \cite{ruffy2018iroko} proposed Iroko, a scalable and modular simulation simulator based on deep reinforcement learning that supported various congestion control algorithms. However, Iroko is still not suitable for large-scale networks, and it cannot adapt to undefined network topologies, where the topology needs to be manually specified, which is not practical in real world scenarios.

\subsubsection{Distributed Congestion Control}
Compared with centralized schemes, distributed schemes decentralize the decision-making authorities and focus more on end-to-end congestion control, concentrating the collaborative algorithm design on distributed network devices and hosts. Majidi et al. used deep learning to improve the processing capability of switches, and to separate elephant flows and mice flows through dual-coupled queues to meet different FCTs \cite{majidi2020dc}. In addition, the ECN threshold of each queue was dynamically tuned to absorb micro-bursts. Nie et al. \cite{nie2019dynamic} proposed a TCP-RL system based on reinforcement learning, which used different learning processes and congestion control strategies for long and short flows to reduce RTT and maximize the overall network throughput. The scheme has been deployed to one of the worldwide top search engines for many years. Whereas, this approach was only evaluated and compared in a static network condition rather than a dynamic DCN.

\subsubsection{Discussion and Insights}
A comparative analysis of some typical representative research works is summarized in Table \ref{Congestion_Control}, and the Table \ref{Assessment of Congestion Control Schemes based on REBEL-3S} shows the evaluation results of each solution according to REBEL-3S. In the light of investigation and evaluation results, we summarize several key problems that could hinder the improvement and implementation of congestion control schemes, as below.

\begin{itemize}
    \item \textbf{Latency.} According to reports provided by cloud service providers \cite{pu2015low}, slight service delays can cause a dramatic drop in user experience, resulting in significant revenue loss. Therefore, ML-based solutions should maximize the user experience by speeding up convergence speed (e.g., using asynchronous components, distributed solutions) and shortening latency as much as possible.
    \item \textbf{Stability of ML algorithms.} The flexibility of ML algorithms is a double-edged sword. Despite its ability to achieve good learning for network fluctuations, it may also become a potential incentive to make the solutions not robust. Furthermore, the instability of ML algorithms may deteriorate the network fluctuations \cite{tokic2011value}.
    \item \textbf{Scheme evaluation.} The experimental evaluations of most existing schemes are based on simulations lacking of verification in real network environment and the network scale is relatively small with simple topologies, which makes the experimental results less convincing.
    \item \textbf{Micro-burst tolerance.} Micro-burst is a common traffic pattern in modern data centers, which can exacerbate the problem of network congestion \cite{jeyakumar2014millions, phanishayee2008measurement, uyeda2011efficiently, shan2017improving}. Reasonable absorption of micro-burst traffic can effectively improve the overall robustness of the network, and performs better than adjusting the congestion window. Unfortunately, there are few ML-based congestion control schemes considering the mitigation of micro-burst, which is believed to be a valuable research topic and a good research opportunity \cite{zou2020flow}.
\end{itemize}

All in all, the current research in this field is still quite limited, and the modeling, algorithm design, experimental method and applications are still in primary research phase, and there is still considerable research value in applying ML to the congestion control.

% Congestion Control
\begin{sidewaystable*}[thp]
    \caption{Research Progress of Data Center Network Intelligence: Congestion Control} % title name of the table
    \label{Congestion_Control}
    \centering % centering table
    \begin{threeparttable}
        \resizebox{ \linewidth}{!}{
            \begin{tabular}{p{4cm} p{1.5cm}<{\centering} p{2.5cm} p{5.3cm} p{3.8cm} p{4.5cm} p{4cm} p{3.5cm} p{3cm}<{\centering}} % creating 8 columns
                \hline\hline % inserting double-line
                \textbf{Ref} & \textbf{Category \tnote{1}}                                                                                                                                                                                                 & \textbf{ML Category \& Model Adopted} & \textbf{Features} & \textbf{Data Source} & \textbf{Feature Selection} & \textbf{Additional Constraints} & \textbf{Estimation Function} & \textbf{Experimental comparison subjects} \\
                \specialrule{0em}{1pt}{1pt}
                \hline % inserts single-line
                \specialrule{0em}{1pt}{1pt}
                % Entering 1st row 
                Jin et al. \cite{jin2018congestion}
                             & C
                             & RL, Q-learning and Sarsa
                             & Applying improved classical reinforcement learning algorithms to software defined data center networks
                             & Simulated data
                             & Flow information, link information, etc.
                             & Link utilization
                             & Occupied bandwidth, average link utilization
                             & \CheckmarkBold  |  \CheckmarkBold  |  \XSolidBrush
                \\
                \specialrule{0em}{1pt}{1pt}
                \hline % inserts single-line
                \specialrule{0em}{1pt}{1pt}
                % Entering 2st row 
                Liao et al. \cite{liao2020fast,liao2019deep}
                             & D
                             & DRL, Q-learning
                             & A low latency and fast converging congestion avoidance strategy
                             & Microsoft Research Cambridge Trace
                             & Average node latency, data block read rate, etc.
                             & None
                             & Read/write latency, convergence speed, etc.
                             & \CheckmarkBold  |  \CheckmarkBold  |  \CheckmarkBold
                \\
                \specialrule{0em}{1pt}{1pt}
                \hline % inserts single-line
                \specialrule{0em}{1pt}{1pt}
                % Entering 3st row 
                Liu et al. \cite{liu2019learning,liu2018learning}
                             & D
                             & DRL, DQN
                             & Reduce service latency with data placement policies
                             & MSR Cambridge Traces \cite{narayanan2008write}
                             & End-to-end node information, read/write latency, subsequent analysis latency, etc.
                             & None (As listed in Table \ref{Resource_Management_1}, the additional constraint for this article is network latency)
                             & Average data read latency, etc.
                             & \CheckmarkBold  |  \CheckmarkBold  |  \CheckmarkBold
                \\
                \specialrule{0em}{1pt}{1pt}
                \hline % inserts single-line
                \specialrule{0em}{1pt}{1pt}
                % Entering 4st row 
                Majidi et al. \cite{majidi2020dc}
                             & D
                             & DL, DNN
                             & Dual queue independent dynamic threshold control based on flow classification for low latency and high throughput
                             & Simulated data
                             & Source IP, destination IP, source port, destination port, protocol information, etc.
                             & High throughput and low losses
                             & FCT, etc.
                             & \CheckmarkBold  |  \CheckmarkBold  |  \XSolidBrush
                \\
                \specialrule{0em}{1pt}{1pt}
                \hline % inserts single-line
                \specialrule{0em}{1pt}{1pt}
                % Entering 5st row 
                Nie \cite{nie2019dynamic}
                             & D
                             & RL, UCB
                             & Presented a TCP-RL system that dynamically configures information flows suitable for short flows via group-based RL and dynamically configures CC schemes suitable for long flows via deep RL
                             & Baidu's Online Production Data Center
                             & Transmission time, throughput, loss rate, RTT, etc.
                             & Maximizing throughput and minimizing RTT
                             & Response time, throughput, RTT, etc.
                             & \CheckmarkBold  |  \CheckmarkBold  |  \CheckmarkBold
                \\
                \specialrule{0em}{1pt}{1pt}
                \hline % inserts single-line
                \specialrule{0em}{1pt}{1pt}
                % Entering 6st row 
                Ruffy et al. \cite{ruffy2018iroko}
                             & C
                             & DRL, DDPG etc.
                             & A simulation simulator supporting different congestion control algorithms
                             & Simulated data
                             & Switch buffer occupancy, interface utilization, etc.
                             & None
                             & Bandwidth, etc.
                             & \CheckmarkBold  |  \CheckmarkBold  |  \CheckmarkBold
                \\
                \specialrule{0em}{1pt}{1pt}
                \hline % inserts single-line
                \specialrule{0em}{1pt}{1pt}
                % Entering 7st row 
                Thiruvenkatam et al. \cite{thiruvenkatam2020optimizing}
                             & D
                             & UL, K-means
                             & Introduced two congestion control mechanisms to improve the transport protocol
                             & Simulated data
                             & Buffer size, dropped packets, etc.
                             & None
                             & Input throughput, etc.
                             & \CheckmarkBold  |  \CheckmarkBold  |  \XSolidBrush
                \\
                \specialrule{0em}{1pt}{1pt}
                \hline % inserts single-line
                \specialrule{0em}{1pt}{1pt}
                % Entering 8st row 
                Sun et al. \cite{sun2020qos,sun2020smartfct}
                             & C
                             & DRL, DDPG
                             & The DRL algorithm was used to improve power efficiency and ensure FCT. Moreover, the link margin ratio was adjusted according to the policy generated by DRL agent to deal with the congestion caused by unpredicted bursts
                             & Wikipedia trace files \cite{urdaneta2009wikipedia}
                             & Number of switch ports, port power, incoming and outgoing flow rates, etc.
                             & QoS and energy consumption
                             & FCT, etc.
                             & \CheckmarkBold  |  \CheckmarkBold  |  \XSolidBrush
                \\
                \specialrule{0em}{1pt}{1pt}
                \hline % inserts single-line
                \specialrule{0em}{1pt}{1pt}
                % Entering 9st row 
                Rastegarfar et al. \cite{rastegarfar2016tcp}
                             & C
                             & DL, DNN
                             & Congestion control through flow classification accuracy and optical bandwidth aggregation
                             & Simulated data
                             & 5-tuple, packet size, etc.
                             & None
                             & Throughput, etc.
                             & \CheckmarkBold  |  \CheckmarkBold  |  \XSolidBrush
                \\
                \specialrule{0em}{1pt}{1pt}
                \hline % inserts single-line
                \specialrule{0em}{1pt}{1pt}
                % Entering 10st row 
                Xiao et al. \cite{xiao2019tcp}
                             & D
                             & DRL, CNN/LSTM
                             & Applied DRL to solve congestion control problems
                             & Simulated data
                             & congestion window size, RTT and inter-arrival time of ACKs
                             & None
                             & Average throughput, average RTT, etc.
                             & \CheckmarkBold  |  \CheckmarkBold  |  \XSolidBrush
                \\
                \hline\hline % inserting double-line
            \end{tabular}
        }
        \begin{tablenotes}
            \footnotesize
            \item[1] For convenience, we use "C" for Centralized Congestion Control, and "D" for Distributed Congestion Control.
        \end{tablenotes}
    \end{threeparttable}
\end{sidewaystable*}

\section{Insights, Challenges and Opportunities}
\label{Insights, Challenges and Opportunities}

\begin{figure*}
    \centering

    \subfigure[From the perspective of research fields of DCN]{
        \includegraphics[width=\linewidth]{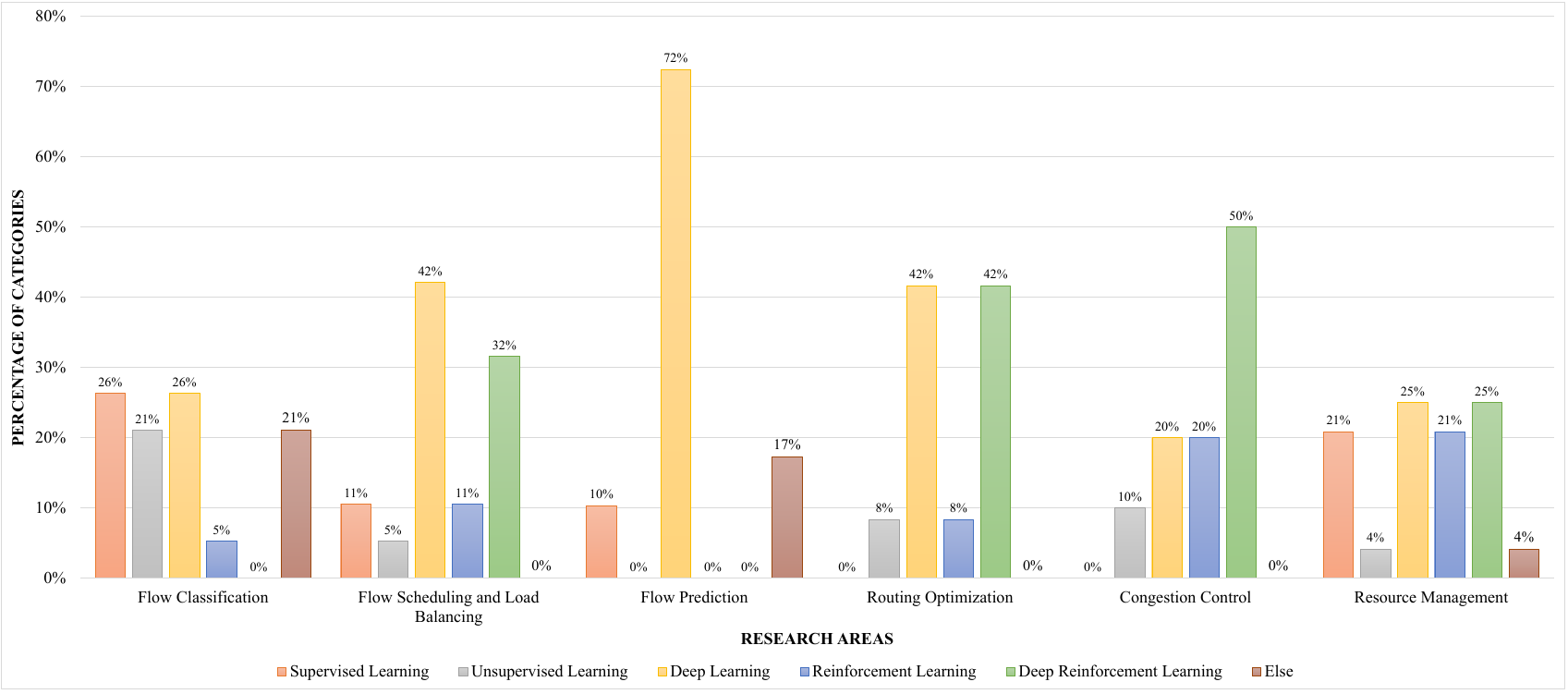}
    }
    \quad
    \subfigure[From the perspective of ML algorithms]{
        \includegraphics[width=\linewidth]{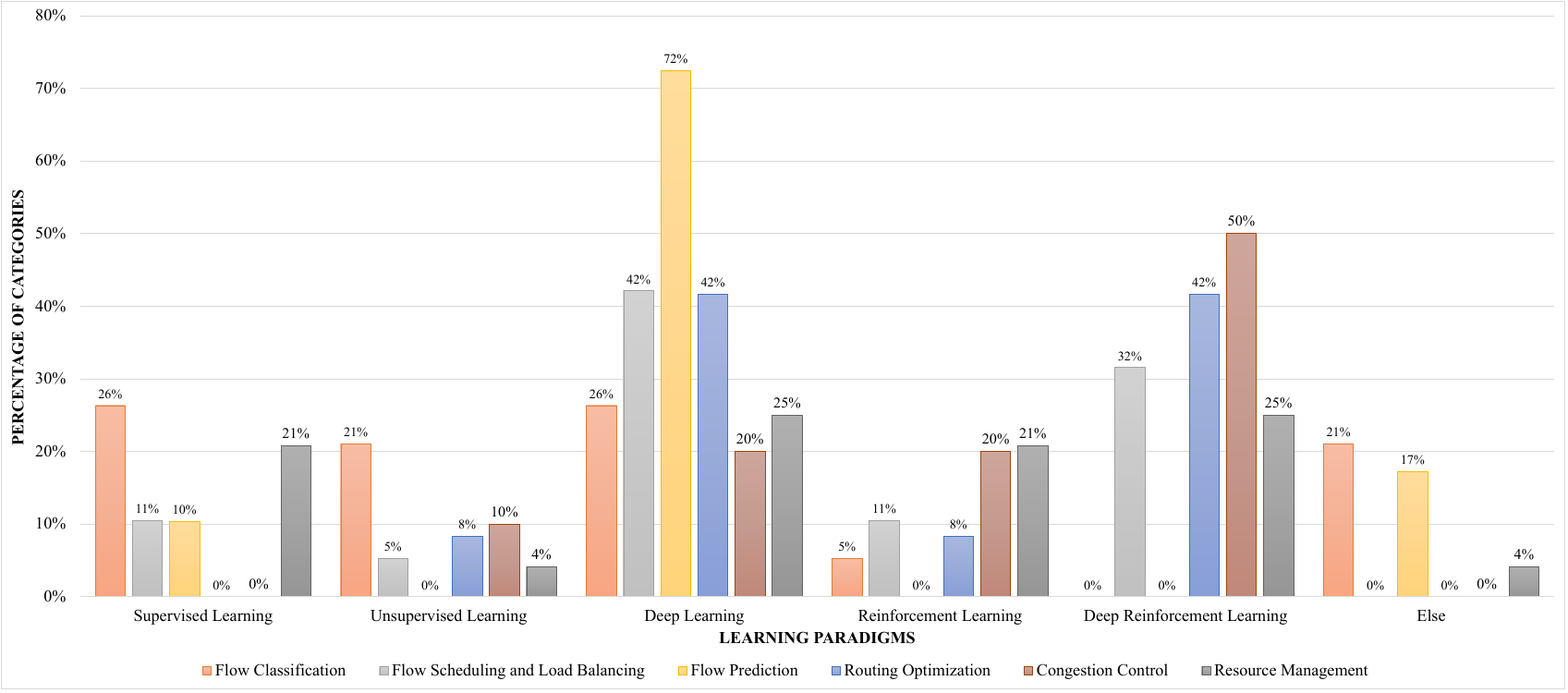}
    }
    \caption{The Current Status of Intelligence In Each Research Area of DCN}
    \label{The_Current_Status_of_Intelligence_In_Each_Research_Area_of_DCN}
    \vspace{-0.3cm}
\end{figure*}

% Heatmap on Cartesian of  various research fields in accordance with REBEL-3S
\begin{figure}
	\centering
	\includegraphics[width=\linewidth]{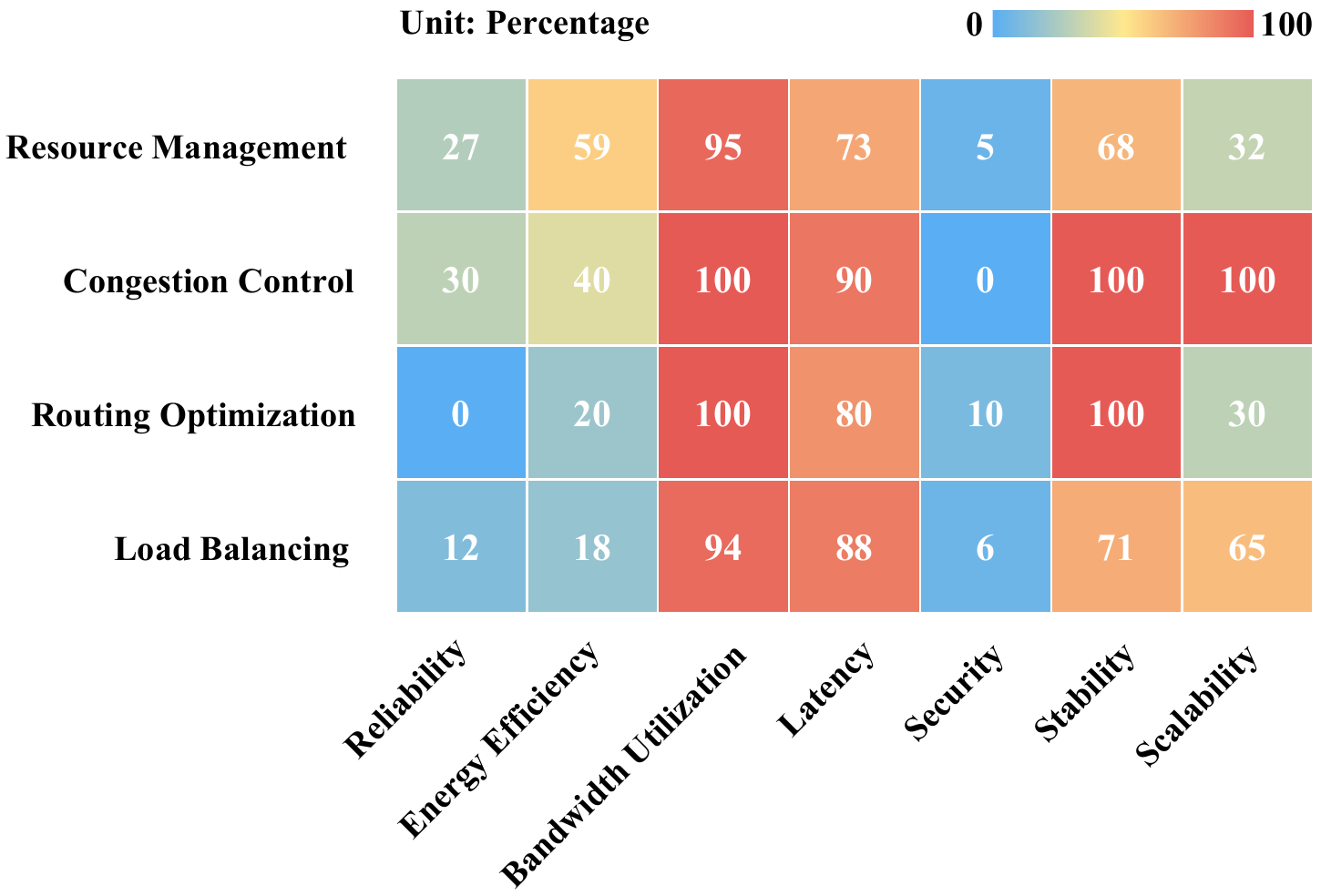}
	\caption{Heatmap on Cartesian of Various Research Fields in Accordance with REBEL-3S}
	\label{Heatmap on Cartesian of  various research fields in accordance with REBEL-3S}
	\vspace{-0.3cm}
\end{figure}

% Challenges_and_Opportunities_of_Intelligent_Data_Center_Networking
\begin{figure}
	\centering
	\includegraphics[width=\linewidth]{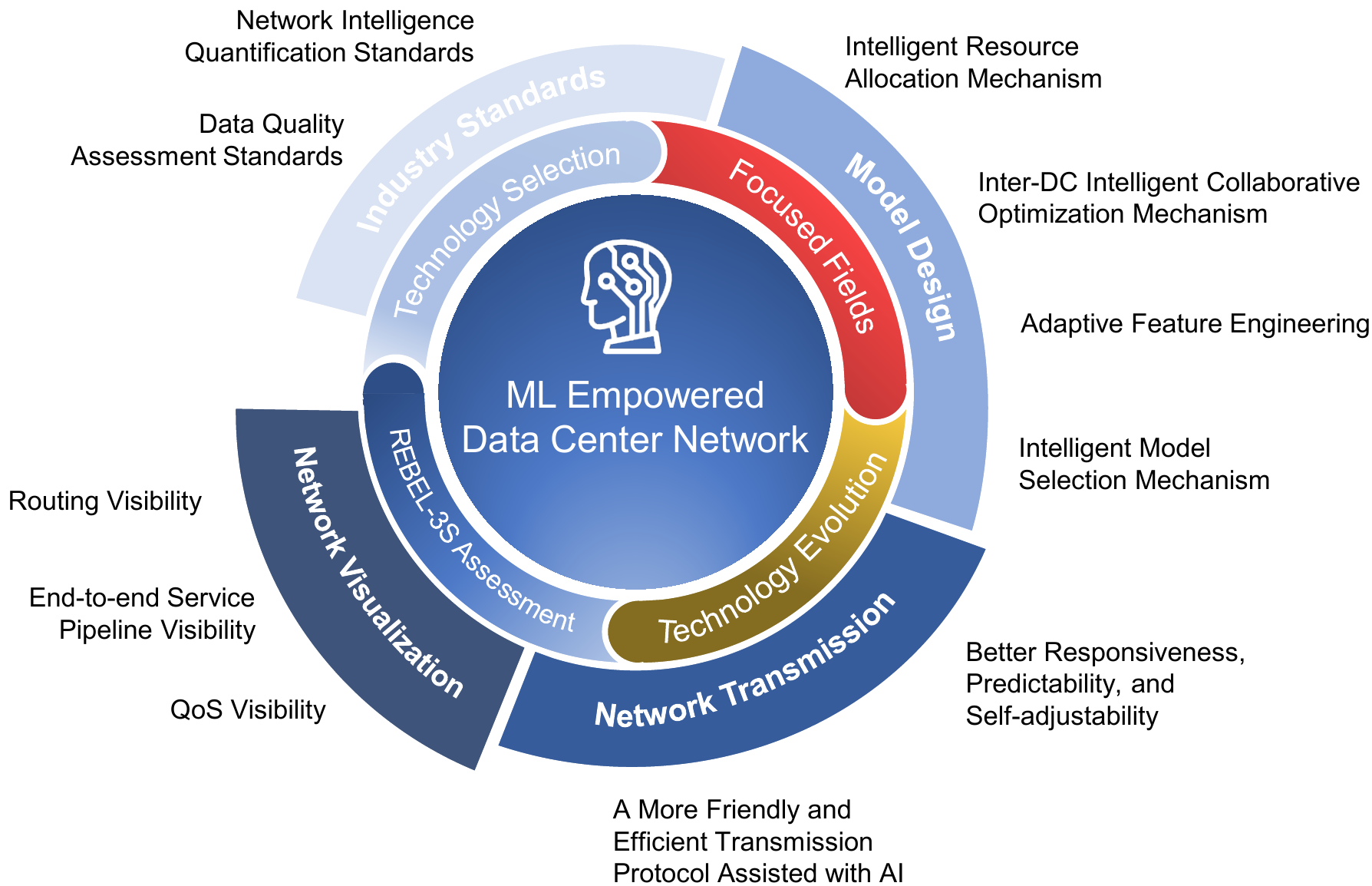}
	\caption{Challenges and Opportunities of Intelligent Data Center Networking}
	\label{Challenges_and_Opportunities_of_Intelligent_Data_Center_Networking}
	\vspace{-0.3cm}
\end{figure}

\begin{table}[htbp]
	\centering
	\caption{Huawei's Five-level Autonomous Driving Network}
	\label{Five-level definition of Huawei's Autonomous Driving Network}
	\begin{spacing}{1.6}
		\begin{threeparttable}
			\resizebox{\linewidth}{!}{
				\begin{tabular}{lllllll}
					\toprule
					\multicolumn{7}{c}{\textbf{Autonomous networks levels \tnote{1}}}
					\\
					\midrule
					\multicolumn{1}{c}{Level Definition}                                                                                                                                                        &
					\multicolumn{1}{p{5.4em}<{\centering}}{\cellcolor[rgb]{ .063,  .176,  .302}\textcolor[rgb]{ 1,  1,  1}
						{    \normalsize \vspace{-0.8em}L0: Manual\newline{\vspace{-0.8em}}Operation \&\newline{}Maintenance}}                                                                                      &
					\multicolumn{1}{p{5.7em}<{\centering}}{\cellcolor[rgb]{ .063,  .176,  .302}\textcolor[rgb]{ 1,  1,  1}{\vspace{-0.8em}L1: Assisted \newline{\vspace{-0.8em}} Operation \&\newline{}Maintenance}
					}                                                                                                                                                                                           &
					\multicolumn{1}{p{5.35em}<{\centering}}{\cellcolor[rgb]{ .063,  .176,  .302}\textcolor[rgb]{ 1,  1,  1}{\vspace{-0.8em}L2: Partial\newline{\vspace{-0.8em}}Autonomous\newline{}Network}}    &
					\multicolumn{1}{p{6.5em}<{\centering}}{\cellcolor[rgb]{ .063,  .176,  .302}\textcolor[rgb]{ 1,  1,  1}{\vspace{-0.8em}L3: Conditional\newline{\vspace{-0.8em}}Autonomous\newline{}Network}} &
					\multicolumn{1}{p{4.9em}<{\centering}}{\cellcolor[rgb]{ .063,  .176,  .302}\textcolor[rgb]{ 1,  1,  1}{\vspace{-0.8em}L4: High\newline{\vspace{-0.8em}}Autonomous\newline{}Network}}        &
					\multicolumn{1}{p{5.3em}<{\centering}}{\cellcolor[rgb]{ .063,  .176,  .302}\textcolor[rgb]{ 1,  1,  1}{\vspace{-0.8em}L5: Full\newline{\vspace{-0.8em}}Autonomous\newline{}Network}}
					\\
					\rowcolor[rgb]{ .851,  .851,  .851} \multicolumn{1}{c}{Execution}                                                                                                                           &
					\multicolumn{1}{c}{\cellcolor[rgb]{ .251,  .447,  .686}P}                                                                                                                                   &
					\multicolumn{1}{c}{\cellcolor[rgb]{ .855,  .886,  .937}P/S}                                                                                                                                 &
					\multicolumn{1}{c}{\cellcolor[rgb]{ .949,  .949,  .949}S}                                                                                                                                   &
					\multicolumn{1}{c}{\cellcolor[rgb]{ .949,  .949,  .949}S}                                                                                                                                   &
					\multicolumn{1}{c}{\cellcolor[rgb]{ .949,  .949,  .949}S}                                                                                                                                   &
					\multicolumn{1}{c}{\cellcolor[rgb]{ .949,  .949,  .949}S}
					\\
					\rowcolor[rgb]{ .851,  .851,  .851} \multicolumn{1}{c}{Awareness}                                                                                                                           &
					\multicolumn{1}{c}{\cellcolor[rgb]{ .251,  .447,  .686}P}                                                                                                                                   &
					\multicolumn{1}{c}{\cellcolor[rgb]{ .251,  .447,  .686}P}                                                                                                                                   &
					\multicolumn{1}{c}{\cellcolor[rgb]{ .855,  .886,  .937}P/S}                                                                                                                                 &
					\multicolumn{1}{c}{\cellcolor[rgb]{ .949,  .949,  .949}S}                                                                                                                                   &
					\multicolumn{1}{c}{\cellcolor[rgb]{ .949,  .949,  .949}S}                                                                                                                                   &
					\multicolumn{1}{c}{\cellcolor[rgb]{ .949,  .949,  .949}S}
					\\
					\rowcolor[rgb]{ .851,  .851,  .851} \multicolumn{1}{c}{Analysis}                                                                                                                            &
					\multicolumn{1}{c}{\cellcolor[rgb]{ .251,  .447,  .686}P}                                                                                                                                   &
					\multicolumn{1}{c}{\cellcolor[rgb]{ .251,  .447,  .686}P}                                                                                                                                   &
					\multicolumn{1}{c}{\cellcolor[rgb]{ .251,  .447,  .686}P}                                                                                                                                   &
					\multicolumn{1}{c}{\cellcolor[rgb]{ .855,  .886,  .937}P/S}                                                                                                                                 &
					\multicolumn{1}{c}{\cellcolor[rgb]{ .949,  .949,  .949}S}                                                                                                                                   &
					\multicolumn{1}{c}{\cellcolor[rgb]{ .949,  .949,  .949}S}
					\\
					\rowcolor[rgb]{ .851,  .851,  .851} \multicolumn{1}{c}{Decision}                                                                                                                            &
					\multicolumn{1}{c}{\cellcolor[rgb]{ .251,  .447,  .686}P}                                                                                                                                   &
					\multicolumn{1}{c}{\cellcolor[rgb]{ .251,  .447,  .686}P}                                                                                                                                   &
					\multicolumn{1}{c}{\cellcolor[rgb]{ .251,  .447,  .686}P}                                                                                                                                   &
					\multicolumn{1}{c}{\cellcolor[rgb]{ .855,  .886,  .937}P/S}                                                                                                                                 &
					\multicolumn{1}{c}{\cellcolor[rgb]{ .949,  .949,  .949}S}                                                                                                                                   &
					\multicolumn{1}{c}{\cellcolor[rgb]{ .949,  .949,  .949}S}
					\\
					\rowcolor[rgb]{ .851,  .851,  .851} \multicolumn{1}{c}{Intent/\newline{}Experience}                                                                                                         &
					\multicolumn{1}{c}{\cellcolor[rgb]{ .251,  .447,  .686}P}                                                                                                                                   &
					\multicolumn{1}{c}{\cellcolor[rgb]{ .251,  .447,  .686}P}                                                                                                                                   &
					\multicolumn{1}{c}{\cellcolor[rgb]{ .251,  .447,  .686}P}                                                                                                                                   &
					\multicolumn{1}{c}{\cellcolor[rgb]{ .251,  .447,  .686}P}                                                                                                                                   &
					\multicolumn{1}{c}{\cellcolor[rgb]{ .855,  .886,  .937}P/S}                                                                                                                                 &
					\multicolumn{1}{c}{\cellcolor[rgb]{ .949,  .949,  .949}S}
					\\
					\rowcolor[rgb]{ .851,  .851,  .851} \multicolumn{1}{c}{Applicability}                                                                                                                       &
					\multicolumn{1}{c | }{\cellcolor[rgb]{ .19,  .33,  .59}\textcolor[rgb]{ 1,  1,  1}{N/A}}                                                                                                    &
					\multicolumn{4}{c | }{\cellcolor[rgb]{ .19,  .33,  .59}\textcolor[rgb]{ 1,  1,  1}{Select scenarios}}                                                                                       &
					\multicolumn{1}{c}{\cellcolor[rgb]{ .19,  .33,  .59}\textcolor[rgb]{ 1,  1,  1}{All scenarios}}
					\\
					\multicolumn{7}{l}{\textbf{P:}Personnel  \textbf{S:}Systems}
					\\
					\bottomrule
				\end{tabular}
			}
			\begin{tablenotes}
				\footnotesize
				\item[1] Table modified based on \emph{Huawei ADN Solution White Paper}.
			\end{tablenotes}
		\end{threeparttable}
	\end{spacing}
\end{table}%

Through systematic research and analysis, we found that ML has been gradually introduced and applied to various fields of data center network, and has made certain achievements. However, the current researches are still in its infancy and need to be further improved in various areas. The survey \cite{noauthor_2020_nodate} by the Uptime Institute in 2020 confirms our view, stating that ML will not take over data center operations and maintenance at this time. In order to further figure out the current progress of ML application in DCN, in this paper we investigate and summarize the popularity of different ML technologies in different DCN fields from different perspectives, as shown in Figure \ref{The_Current_Status_of_Intelligence_In_Each_Research_Area_of_DCN}. Moreover, based on the statistics of the existing work, we make a further analysis from the aspects of ML technology selection, focuses of DCN fields, and REBEL-3S assessment, and provide some more in-depth insights.

\begin{enumerate}
    \item \textbf{Technology Selection:} ML has been carried out in a series of work in various research areas of DCN. Deep learning has gained the favor of researchers because of its good comprehensive ability, accounting for over 50\% of all solutions. Supervised learning and deep reinforcement learning are ranked second and third, respectively. According to our current research, the lack of universality \cite{zhang2018study,whiteson2011protecting,raghu2018can,lanctot2017unified} and reproducibility \cite{henderson2018deep} are the important reasons why reinforcement learning only ranks third up to now. As for the experimental verification of these schemes, over 35\% of schemes were conducted based on simulated data other than real-world data, which lacks convincing results to prove their effectiveness in real-world environments.
    \item \textbf{Focuses of DCN Fields:} The application progress of ML in different fields of DCN also varies. Currently, the researchers mainly focus on flow prediction, resource management, flow classification, flow scheduling, and load balancing, but pay less attention to route optimization, and congestion control.
    \item \textbf{REBEL-3S Assessment:} In order to more accurately assess the current research status of data center network intelligence, we further analyzed the existing research work according to the proposed REBEL-3S assessment criteria. Besides, we summarize the research progress of ML-based intelligent DCN and draw a vivid Heatmap on Cartesian, as shown in Figure \ref{Heatmap on Cartesian of  various research fields in accordance with REBEL-3S}, where the "27" in the top-left corner, for example, represents that 27\% of the research work in the field of resource management considers RELIABILITY. Clearly, it can be seen that most of the research work has considered bandwidth utilization and stability, where the values of both two columns are above 65\%. Comparatively, most of the research work lacks attention to security and reliability, and the results show that most of the values of these two columns remain below 30\%. On the other hand, Figure \ref{Heatmap on Cartesian of  various research fields in accordance with REBEL-3S} also reflects that the research work in the fields of resource management, flow scheduling and load balancing considers more dimensions of REBEL-3S, and the solutions are more mature.
\end{enumerate}

Overall, most of the existing work still has some unresolved issues, and there are still many opportunities to explore and also a lot of room to improve the level of intelligence. Our view is that the future data center network should be endogenously embedded with intelligence. It is admitted that up to now ML has gained much popularity in various industries, especially in the field of data center network, however, currently its role is more like a tool or module grafted in the system. Whereas, we insist that the intelligence of the future data center network should be an intrinsic natural attribute.

In addition, the network intelligence have been discussed for a long time, but how to define and quantify network intelligence still remains not standardized, and there is not a recognized measurement criterion as well. Huawei has proposed a grading scheme of intelligent networks \cite{autonomous}, as shown in Table \ref{Five-level definition of Huawei's Autonomous Driving Network}, which defines six levels of intelligence, ranging from L0 to L5. The L0 intelligence has the ability of auxiliary monitoring, and the execution of all dynamic tasks still depends on manual operations, while the highest L5 intelligence realizes a fully autonomous network with full life cycle closed loop automation capabilities across multiple services and domains. However, this grading scheme still does not provide a formulaic quantitative method to directly and accurately quantify the intelligence level of a network. Therefore, there is still a strong need to further explore a specific quantitative formula for grading the degree of intelligence similar to the Shannon formula.

Finally, before concluding this paper, we will further discuss the challenges and opportunities of data center network intelligence from four aspects: industry standards, model design, network transmission and network visualization, as shown in Figure \ref{Challenges_and_Opportunities_of_Intelligent_Data_Center_Networking}.

\subsection{Industry Standards}

\subsubsection{Network Intelligence Quantification Standards}
As aforementioned, because the research on ML-based intelligent data center is still in the initial stage, both the academia and industry have not formulated a specific quantitative standard to assess the network intelligence level. Although some leading high-tech companies (e.g. Huawei) have proposed some directive principles for grading network intelligence levels, which only defines the characteristics and capabilities of the network with different intelligence levels,  this is far from being adequate.
The ultimate goal is to design a mathematical formula like intelligence quantification method with fairness and accuracy, though there is a long way to go.

\subsubsection{Data Quality Assessment Standards}
\label{Data Quality Assessment Standards}
The quality of the source data includes authenticity, validity, diversity, and timeliness. Simulated data lack convincingness, scenario-specific generated data lack universal validity, data containing only a few feature information is challenging to improve the accuracy of predictions, and antiquated historical data lose timeliness having little value. Some existing solutions did not provide any information about the data source, making it difficult to examine the quality of data sets and the validity of experiments. It is necessary to call on researchers to develop a data quality assessment standard as soon as possible, where a quantifiable data quality assessment standard will greatly help enhance the convincingness of experimental results and advance the network intelligence process to some extent.

\subsection{Model Design}
\subsubsection{Intelligent Resource Allocation Mechanism}
Data center networks need to intelligently perceive scenarios and services, and reasonably consider the lifecycle of resource management, i.e., resource prediction, allocation, utilization, integration, and recovery, under the security conditions. However, most the existing solutions mainly focus on the resource prediction, allocation and utilization optimization, and there is little research on the resource fragment integration and resource recovery.

\subsubsection{Inter-DC Intelligent Collaborative Optimization Mechanism}
The inter-DC network optimization is also a very important but more complex research topic, where the optimization usually requires close collaborations among multiple data centers. Thereby, how to achieve efficient collaboration among different intelligent models of different data centers has become a big challenge. Ideally, all separate models can be globally trained based on a complete set of all DC's data, however, normally local data cannot be transferred freely across data centers due to privacy and bandwidth overhead issues. Hence, it will be a good research opportunity to explore efficient methods to achieve an efficient collaboration of inter-DC intelligent models on the premise of ensuring data privacy and security.

\subsubsection{Adaptive Feature Engineering}
Feature engineering largely affects the ultimate effect of machine learning models. Usually, the feature engineering in ML models is specially designed for a single problem in a specific scenario. However, the richness of data center network layers makes the data collected at each layer vary greatly, and the diversity of services also makes the corresponding feature selection different. How to make feature engineering adaptive to network scenarios and service types under the above complex environment is a key challenge for feature engineering in DCN.

\subsubsection{Intelligent Model Selection Mechanism}
Without doubt, there is no one universal learning model that works for all scenarios, and every model has certain limitations in different scenarios. The highly dynamic nature of the network environment, the diversity of service requirements, the heterogeneity of network data, and the inconsistency of optimization goals make it extremely difficult and time-consuming to select the most suitable machine learning model. This is also one of the key pain points of the application of artificial intelligence in data center networks.

\subsection{Network Transmission}
As the data center service scenarios become more and more complex, the network scale becomes larger and larger, and the requirements for user experience and service quality become higher and higher, the traditional communication protocols have already not been qualified to cope with these challenges. Thus, the data center network inevitably requires more efficient and intelligent communication protocols to ensure fast convergence, high bandwidth, low latency and no packet loss for network transmissions. However, the new communication protocols proposed in recent years also fail to meet data center requirements of heterogeneous scenarios \cite{judd2015attaining} and have compatibility issues with legacy protocols \cite{he2016ac,cronkite2016virtualized}. There is no doubt that artificial intelligence can help network protocols achieve better responsiveness, predictability, and self-adjusting ability. However, there is still little research on how to achieve a more friendly and efficient transmission protocol assisted with ML, which is an opportunity for future research.

\subsection{Network Visualization}
Due to the rapid growth of data center networks, the network size has expanded dramatically and the amount of network data generated has increased greatly. As a result, the burden of network supervision is getting heavier. A proven way to accurately monitor and control from the massive amount of data is network visualization. Network visualization is a comprehensive and concise display of network data by means of graphics, and its ability to reduce the burden of traditional network monitoring. Network visualization can help the O\&M personnels accurately perceive the network by explicitly presenting the real-time status of the network to them. Currently, the data center network still has the following three aspects of invisibility, which leads to inefficiency of network O\&M and optimization.
\begin{itemize}
    \item \textbf{Routing invisible:} Invisible routing makes the transmission changes cannot be reproduced and the changing process cannot be backtracked. This often leads to a tough situation, that is, the user reported one network fault, but when the O\&M personnel starts to locate the fault, the fault disappears again, and there is no historical information to query. As a result, the cause of the fault cannot be diagnosed.
    \item \textbf{End-to-end service pipeline invisible:} This leads to the inability to see the actual forwarding path corresponding to the service pipeline, as well as the performance of the forwarding path. As a result, after the network failure occurs, we can only locate the failure hop by hop, which is time-consuming and laborious.
    \item \textbf{QoS invisible:} The service quality is not visible, resulting in the user experience cannot be perceived. The traditional network management tools usually can only provide the performance data of network, but cannot exhibit the quality of the service contents carried by the network. In other words, the network performance and service quality are separated without any correlations, resulting in low efficiency of fault location.
\end{itemize}

Overall, the research on applying ML to achieve and intelligent data center networking is still at an early stage. There are always many opportunities to further explore the potential and value of applying ML technologies in various fields of data center networks. It can conclude that the network intelligence will inevitably become the future trend of data center network development. In the foreseeable future, ML-based intelligent networking will become the core research direction of the cloud computing, driving the data center network from SDN-enabled automatic network to ML-driven intelligent network.

\section{Conclusion}
\label{Conclusion}
As the core infrastructure, data center provides a strong platform support for cloud computing, and so on. Nevertheless, the rapid growth of its network scale leads to great challenges in network optimization. Fortunately, artificial intelligence provides a promising way to deal these challenges, and it has been successfully employed in various fields of DCNs. Up to now, there have been numerous literature on ML-assisted intelligent data center networking. However, to the best of our knowledge, there is a lack of systematic investigations into these research works. To this end, in this survey paper, we comprehensively review the representative research works with in-depth analysis and discussions from various perspectives including flow prediction, flow classification, load balancing, resource management, routing optimization, and congestion control. Notably, to better assess the existing works we creatively propose the REBEL-3S quality assessment scheme. Finally, we thoroughly explore the challenges existed in current research and opportunities for future research from various aspects together with our key insights. To sum up, the research on the application of artificial intelligence in data center networks is still in its infancy, but it has aroused the attention of more and more scholars and researchers, and has achieved preliminary research results in many fields. However, there are still many problems and deficiencies in the current research, which remain to be further studied.

% \section*{Acknowledgment}

\ifCLASSOPTIONcaptionsoff
    \newpage
\fi

\begin{spacing}{1}
    \footnotesize
    \bibliographystyle{IEEEtran.bst}
    \bibliography{./main}
\end{spacing}

% \bibliography{./reference_list/references2.bib}
\end{document}